\documentclass[pra,twocolumn,nofootinbib,floatfix,10pt]{revtex4-2}

\usepackage{amsmath}
\usepackage{amssymb}
\usepackage{wasysym}
\usepackage{graphicx}
\usepackage{color,soul}
\usepackage{physics}
\usepackage{siunitx}
\usepackage{dsfont}
\usepackage{float}
\usepackage{xcolor}
\usepackage[english]{babel}
\usepackage{blindtext}
\usepackage[english,nomargin,inline,marginclue,draft]{fixme}
\pdfpageheight\paperheight
\pdfpagewidth\paperwidth

\usepackage[colorlinks,linkcolor=blue,anchorcolor=blue,citecolor=blue,urlcolor=blue]{hyperref}

\fxusetheme{colorsig}
\FXRegisterAuthor{cg}{acg}{CG}  
\FXRegisterAuthor{th}{ath}{\color{blue}TH}  
\FXRegisterAuthor{ib}{aib}{\color{red}IB} 
\FXRegisterAuthor{sh}{ash}{\color{cyan}SH} 
\FXRegisterAuthor{db}{adb}{\color{green}DB} 
\FXRegisterAuthor{ps}{aps}{PS}
\makeatletter
\renewcommand*\FXLayoutInline[3]{%
  {\@fxuseface{inline}\ignorespaces{\color{fx#1}[#3: #2]}}}
\makeatother

\long\def\symbolfootnote[#1]#2{\begingroup%
\def\thefootnote{\fnsymbol{footnote}}\footnotetext[#1]{#2}\endgroup}

\def\nobreakbefore{%
  \relax\ifvmode\else
    \ifhmode
      \ifdim\lastskip > 0pt\relax
        \unskip\nobreakspace
      \else 
        \nobreakspace
      \fi
    \fi
  \fi
}
\let\oldcite\cite
\renewcommand\cite{\nobreakbefore\oldcite}

\begin{document}
\title{Compressive Spectrum Sensing via Spectral Multiplexing in Rydberg Atomic Receiver}

\author{Jun-Rong Chen$^{1,\textcolor{blue}{\star}}$}
\author{Yi-Ming Yin$^{2,3,\textcolor{blue}{\star}}$}
\author{Le-Bin Chen$^{6,\textcolor{blue}{\star}}$}
\author{Kai Wang$^{4,\textcolor{blue}{\star}}$}
\author{Bang Liu$^{2,3}$}
\author{Li-Hua Zhang$^{2,3}$}
\author{Hao Tian$^{1,\textcolor{blue}{\P}}$}
\author{Ming-Min Zhao$^{6,\textcolor{blue}{\S}}$}
\author{Bin-Bin Wei$^{5,\textcolor{blue}{\ddagger}}$}
\author{Dong-Sheng Ding$^{2,3,\textcolor{blue}{\dagger}}$}

\affiliation{$^1$School of physics, Harbin Institute of Technology, Harbin, Heilongjiang 150001, China.}
\affiliation{$^2$Key Laboratory of Quantum Information, University of Science and Technology of China, Hefei, Anhui 230026, China.}
\affiliation{$^3$Anhui Province Key Laboratory of Quantum Network, University of Science and Technology of China, Hefei, Anhui 230026, China.}
\affiliation{$^4$Electronic Engineering Institute, National University of Defense Technology, Hefei, Anhui 230037, China}
\affiliation{$^5$Institute of System Engineering, Tianjin 300161, China.}
\affiliation{$^6$College of Information Science and Electronic Engineering, Zhejiang University, Hangzhou, Zhejiang 310058, China}

\date{\today}

\symbolfootnote[1]{J.R.C., Y.M.Y., L.B.C. and K.W. contribute equally to this work.}
\symbolfootnote[2]{dds@ustc.edu.cn}
\symbolfootnote[3]{weibb.2009@tsinghua.org.cn}
\symbolfootnote[4]{zmmblack@zju.edu.cn}
\symbolfootnote[5]{tianhao@hit.edu.cn}

\maketitle

\textbf{
    Rydberg‑atomic receivers exhibit exceptional sensitivity yet are fundamentally constrained by the narrow instantaneous bandwidth, limiting their practical deployment in broadband scenarios. Prior approaches typically expand the bandwidth by physically broadening the atomic response, which usually requires auxiliary electromagnetic fields or stringent parameter tuning, thereby increasing overall system complexity. Here, we propose a compressive spectral multiplexing framework implemented in a waveguide-coupled Rydberg atomic receiver using a frequency-modulated local oscillator (FMLO). The FMLO creates multiple parallel sensing channels that collectively constitute a physical compressive sensing matrix, generating multiple narrowband intermediate-frequency replicas of the input signal. Thus, a broadband microwave spectrum is projected onto a set of narrowband atomic responses. It is demonstrated that spectral information spanning a bandwidth of over 640 MHz can be effectively compressed into the intrinsic atomic bandwidth of 126 kHz, achieving a spectrum compression ratio exceeding 1000. Furthermore, these output replicas offer intrinsic measurement redundancy and facilitate signal‑to‑noise ratio enhancement. An approximate 10 dB gain is achieved in the required bit‑energy‑to‑noise‑power‑density ratio for multi-channel communication via maximal‑ratio combining. This approach requires no auxiliary fields or broadband electronics, providing a simple and scalable pathway for chip-scale quantum receivers, latency-critical sensing, and next-generation wireless communications.
}

\section*{Introduction}

    With the rapid advancement of wireless communication technologies, the electromagnetic spectrum has emerged as a critical strategic resource increasingly constrained by congestion and scarcity \cite{torrieri2005principles, 9614463}. Accordingly, broadband identification and real-time monitoring of electromagnetic signals are essential for dynamic spectrum sharing, cognitive radio networks, and multi-band coordination in next-generation communications \cite{cuiRydbergAtomicReceiver2026}. Nevertheless, conventional microwave receivers based on metal antennas, despite their technological maturity, inherently suffer trade-offs among sensitivity, instantaneous bandwidth, and hardware complexity when pursuing broadband coverage. The high cost and power consumption associated with high-sampling-rate analog-to-digital converters, complex filter banks, and swept architectures have driven research into innovative broadband sensing paradigms \cite{10.1145/3570361.3613258, allinsonRydbergReceiversSpace2026, gongRydbergAtomicQuantum2025}. Especially, for spectrally sparse signals, conventional compressive sensing (CS) enables the acquisition of broadband information at sub-Nyquist sampling rates, providing a viable pathway to decouple spectral coverage from measurement bandwidth. However, practical implementations of this approach generally rely on random measurement matrices followed by L1-norm optimization for signal recovery , requiring specialized hardware architectures that remain challenging to realize in practice. A more favorable alternative is a physical system that inherently performs compressive projections, eliminating the need for artificially engineered random matrices and iterative reconstruction algorithms.

    Rydberg‑atom‑based quantum microwave receivers constitute a compelling physical platform for this purpose. They deliver ultrahigh sensitivity, quasi‑continuous spectral coverage extending toward the terahertz band, and intrinsic quantum traceability \cite{holloway2022overview, liuElectricFieldMeasurement2023a, schlossbergerRydbergStatesAlkali2024a, somaweeraRydbergAtomBasedSensors2025, yuanQuantumSensingMicrowave2023}. The introduction of electromagnetically induced transparency (EIT) and Autler-Townes (AT) splitting, combined with microwave local oscillator (LO) and heterodyne detection schemes, has significantly improved the electric field detection sensitivity of Rydberg receivers  \cite{jingAtomicSuperheterodyneReceiver2020, yuanRydbergAtomBasedReceiver2023a}.
    State-of-the-art Rydberg atomic receivers have demonstrated sensitivities in the $\rm{nV\cdot cm^{-1}\cdot Hz^{-1/2}}$ regime, rendering them promising tools for weak-field detection \cite{wangQuantumEnhancedMetrology2026, liuEnhancedMultiparameterMetrology2026, xiaoLowfrequencyWeakElectric2024, liangCavityEnhancedRydbergAtomic2025, venu2025three, schlossbergerFundamentalLinewidthLimit2026, prajapatiSensitivityComparisonTwophoton2023, schlossbergerElectromagneticallyInducedTransparency2026, tu2024approaching, liuContinuousFrequencyMicrowaveHeterodyne2022a, zhangAdvancingThreePhotonExcitedRydberg2025a}. Beyond high sensitivity, the versatility of Rydberg atoms as microwave receivers has been extensively explored. They permit high‑precision characterization of microwave field parameters, including amplitude \cite{jingAtomicSuperheterodyneReceiver2020, yaoUltrawidebandContinuousSpectrum2026, jiangQuantumWeakMeasurement2025}, frequency \cite{chenInstantaneousFrequencyEstimation2024, liuDeepLearningEnhanced2022a, gordonWeakElectricfieldDetection2019}, phase \cite{hanPhasesensitiveRydbergatomInterferometry2025, berwegerClosedloopQuantumInterferometry2023, simonsRydbergAtombasedMixer2019}, polarization \cite{berwegerIndependentRydbergAtom2024, elgeeElectricallySmallRydberg2025, changEffectCouplingBeam2025a, youRFSpectraInduced2024, schlossbergerZeemanresolvedAutlerTownesSplitting2024}, and wavevector orientation \cite{oliverSimultaneousDetectionDemodulation2026, talashilaDeterminingAngleArrival2025, chen2025multi, maoDigitalBeamformingReceiving2024}. These capabilities have facilitated a range of practical applications, such as wireless communication \cite{tongBroadbandFrequencyHopping2026, wen2024rydberg, pengEnhancedGroundSatelliteDirect2026, qiuShortwaveFrequencyhoppingReception2026, xieAtomicMicrowaveSensing2025, knarrSpatiotemporalMultiplexedRydberg2023a, meyerSimultaneousMultibandDemodulation2023}, remote sensing \cite{allinsonRydbergReceiversSpace2026, arumugamRemoteSensingSoil2024}, imaging \cite{wattersonImagingRadarUsing2025}, low-frequency signal acquisition \cite{bangliuHighlySensitiveMeasurement2022, lei2024high, xie2026optical, liuContinuousFrequencyMicrowaveHeterodyne2022a, arumugamStarkModulatedRydberg2025, jinHeterodyneDetectionLowFrequency2025, chandraElectrometryExtremelylowFrequencies2026, hammerlandMHzSubkHzField2026}, radar detection \cite{bohaichukOriginsRydbergAtomElectrometer2022, chenHighResolutionQuantumSensing2025}, etc. Despite these advances, extending Rydberg receivers to broadband operation remains challenging due to their limited instantaneous bandwidth. This bandwidth defines the spectral range over which signals can be simultaneously measured without distortion, yet in practice it is typically limited to only hundreds of kilohertz to several megahertz \cite{jingAtomicSuperheterodyneReceiver2020, yuanRydbergAtomBasedReceiver2023a, lihuazhangUltrawideDualbandRydberg2024, tu2022high}, far below the demands of broadband scenarios.  Existing bandwidth‑extension strategies mainly focus on broadening the effective instantaneous bandwidth of the atomic system itself, including optical field engineering \cite{prajapatiTVVideoGame2022a}, multi-wave mixing \cite{yanMultiDressStateEngineeredRydberg2025, bowenyangHighlySensitiveMicrowave2024}, multi-channel excitation \cite{huImprovementResponseBandwidth2023a}, frequency-comb-assisted techniques \cite{artusio_glimpseIncreasedInstantaneousBandwidth2024, zhang2024floquet, lihuazhangTunableOffresonantRydberg2024, chenInstantaneousFrequencyEstimation2024}, as well as energy-level engineering with Zeeman effect \cite{qimengInstantaneousBandwidthExpansion2025} and Stark shifting \cite{jiaoArbitraryInstantaneousBandwidth2025}. 
    
    While these approaches have successfully extended the equivalent sensing bandwidth to tens of megahertz, they generally entail intricate optical or microwave hardware configurations and increased system cost. The receivers also fundamentally remain single-channel and lack inherent measurement redundancy for improved detection reliability. However, in many practical scenarios, broadband signals occupy sparse or fragmented spectra instead of continuous spectra. Detection of such signals does not require a proportionally large instantaneous bandwidth, but rather the ability to capture the essential spectral features of the signal. This perspective motivates an alternative sensing paradigm, where broadband information is acquired through structured projections rather than direct broadband reception. A broadband sparse spectrum can thus be mapped onto multiple narrowband measurements via parallel mixing channels, effectively compressing the spectral information into the intrinsic bandwidth of the receiver. This approach is particularly well suited for Rydberg atomic systems with inherently limited instantaneous bandwidth, while also enabling redundant measurements that improve the detection reliability.
    
    In this work, we move beyond the prevailing paradigm of broadening the intrinsic atomic response and instead propose a structured spectral multiplexing framework. Implemented on a waveguide-coupled Rydberg receiver, this framework provides a compact and scalable interface for microwave field delivery and leverages the intrinsic multi-heterodyne process of the Rydberg system as a physical realization of compressive measurement. Central to this implementation is the replacement of the conventional single-tone LO with a FMLO, which produces a spectrum of equally spaced, phase-locked frequency components. These components act as parallel down-conversion channels, mapping a broad spectral range into a set of narrowband intermediate-frequency (IF) outputs. This defines a sensing matrix determined by the FMLO spectral envelope. Under this framework, we systematically investigate the 1-sparse and multi-frequency response characteristics of the system and elucidate the underlying mixing mechanisms. Leveraging this multi-heterodyne mapping, we demonstrate a recoverable spectral range exceeding 640 MHz, compressing more than three orders of magnitude of spectral information into the intrinsic atomic bandwidth. Furthermore, we demonstrate that the spectral diversity enables signal-to-noise ratio (SNR) improvement with maximal-ratio combining (MRC) in multi-channel communication experiments, yielding an approximate 10 dB gain in the required equivalent bit-energy-to-noise-power-density ratio $E_b/N_0$. Relying solely on a single readily generated microwave waveform, this approach eliminates the need for auxiliary fields and broadband electronics, enabling a scalable route toward broadband spectrum sensing, chip‑scale quantum receivers, and next‑generation wireless communications.

    \begin{figure*}
        \centering
        \includegraphics[width=1\linewidth]{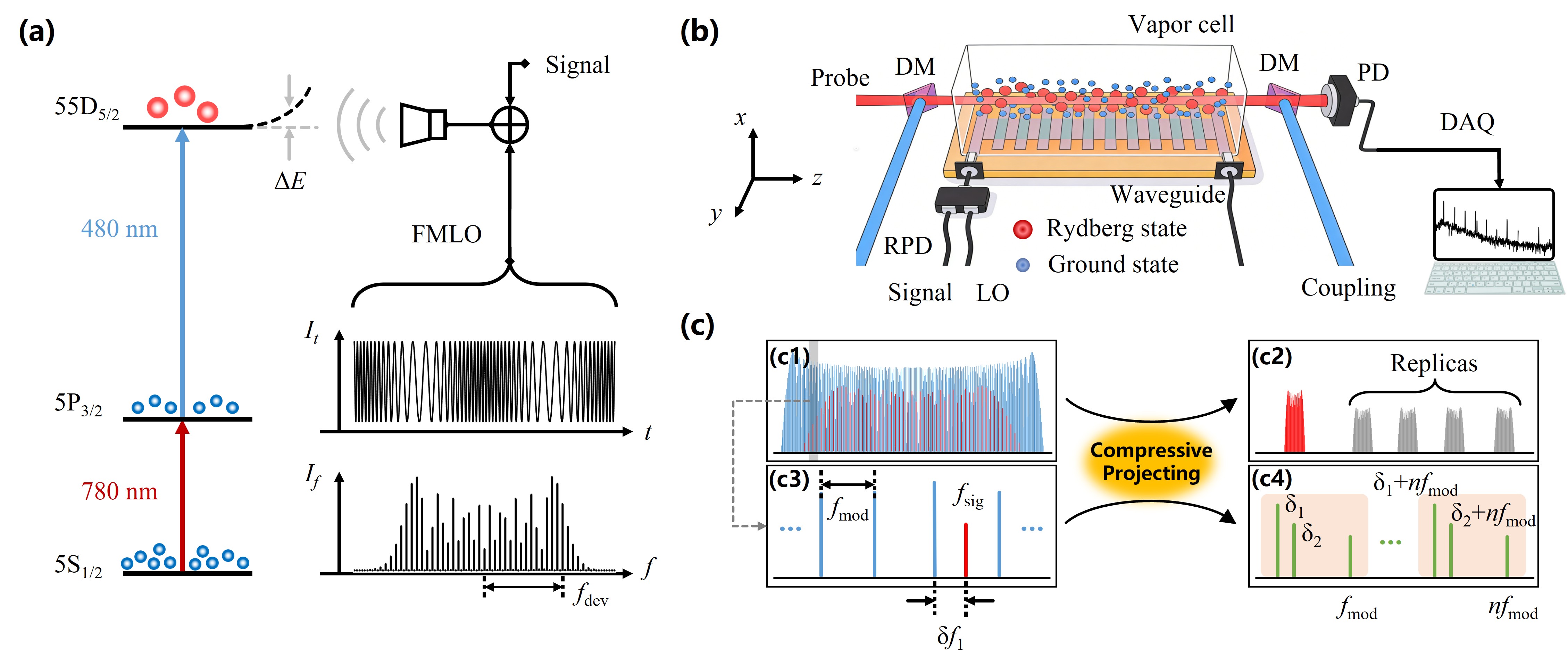}
        \caption{\textbf{Compressive spectral multiplexing via a frequency-modulated local oscillator.} (a) Energy level diagram of $^{85}\rm{Rb}$ Rydberg atoms. The probe laser (780 nm, waist 320 $\rm{\mu m}$) is resonant with the $5\rm{S}_{1/2} \rightarrow 5\rm{P}_{3/2}$ transition, while the coupling laser (480 nm, waist 560 $\rm{\mu m}$) drives the  $5\rm{P}_{3/2} \rightarrow 55\rm{D}_{5/2}$ transition to the Rydberg state. The LO is configured as a frequency-modulated (FM) signal, with its time-domain and frequency-domain representations illustrated in the upper and lower panels on the right, respectively. Under the FMLO, the Rydberg atoms experience AC Stark shifts that mediate the compressive multi-heterodyne mixing with external signal fields. (b) Schematic of the experimental setup. The FMLO and the signal are combined via a resistive power divider (RPD) and coupled into the vapor cell through a waveguide. The probe and coupling lasers propagate in a counter-collinear configuration. A dichroic mirror (DM) is used to separate the probe light, and the response of the Rydberg atoms is read out using a photodetector (PD) and a data acquisition card (DAQ). (c) Illustration of the compressive spectral multiplexing mechanism. (c1) A broadband multi-tone signal (red) overlaps with the FMLO spectrum (blue) and undergoes multi-heterodyne mixing in Rydberg atoms. (c2) The spectrum of the signal is compressed into multiple narrowband IF responses, including a zeroth-order component and its replicas. (c3, c4) Local mixing of a single spectral component with adjacent FMLO tones, illustrating the formation of multiple compressive projections.}
        \label{fig1}
    \end{figure*}

    \begin{figure*}
        \centering
        \includegraphics[width=1\linewidth]{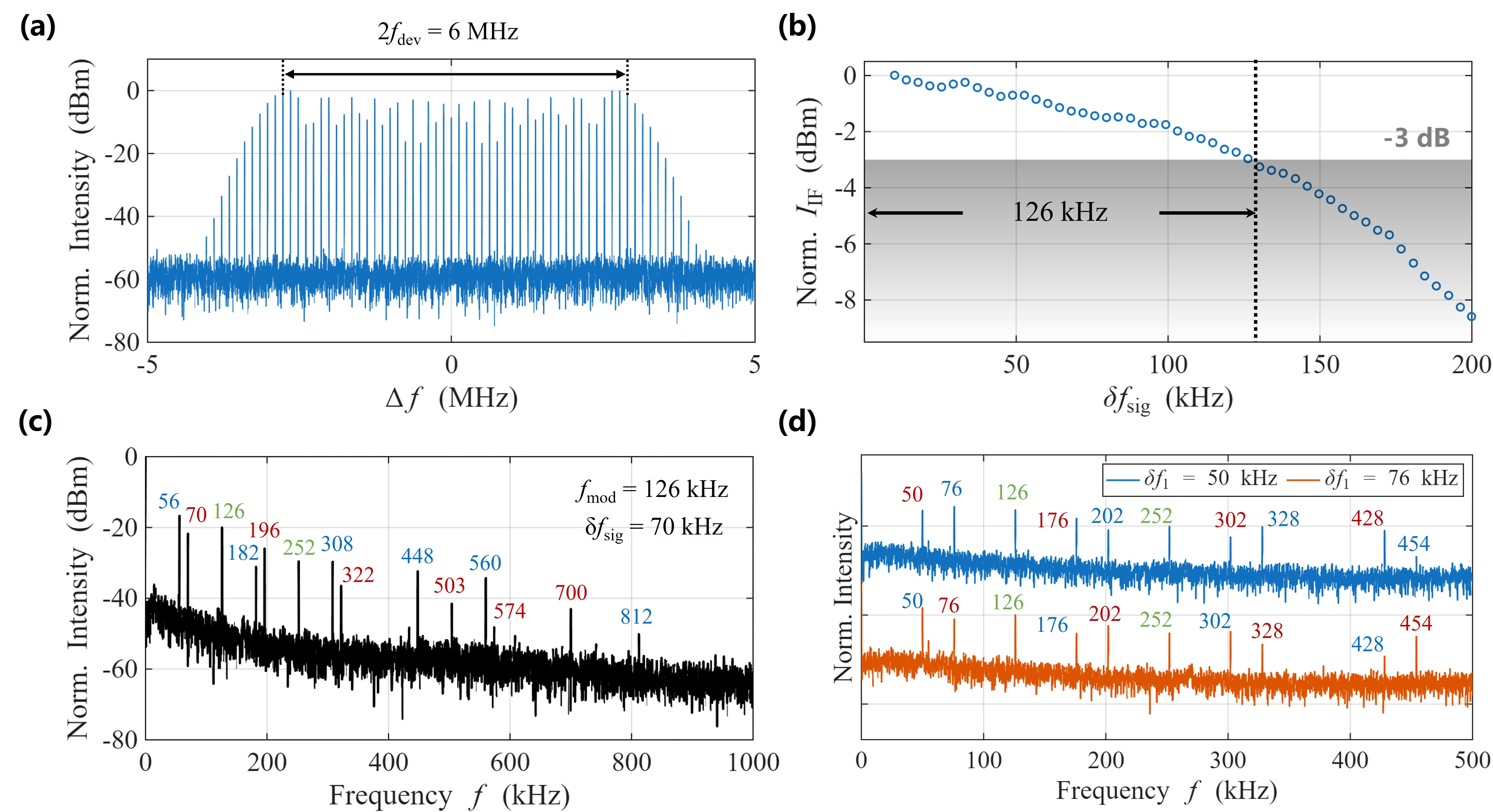}
        \caption{\textbf{Single-frequency response under compressive spectral multiplexing.} (a) Frequency spectrum of the FMLO used in the experiment. The frequency deviation, the center frequency, and the modulation rate are fixed at $f_{\rm{dev}} = 3~\rm{MHz}$, $f_{\rm{c}} = 4.5~\rm{GHz}$, $f_{\rm{mod}} = 126~\rm{kHz}$, respectively. (b) The response amplitude of the Rydberg system under a single-frequency LO at 4.5 GHz is measured as a function of the frequency detuning between the signal field and the LO field ($\delta f_{\rm{sig}} = |f_{\rm{sig}} - f_{\rm{LO}}|$), yielding an instantaneous bandwidth of approximately 126 kHz. (c) Output signal spectrum of the Rydberg atoms when a single-frequency signal with a detuning of $\delta f_{\rm{sig}} = 70~\rm{kHz}$ is applied in addition to the FMLO. The spectrum contains IF components from three families, $nf_{\rm{mod}}+\delta f_{\rm{sig}}$ (red), $nf_{\rm{mod}}-\delta f_{\rm{sig}}$ (blue) and $nf_{\rm{mod}}$ (green), where $n\in \mathbb{N}$, corresponding to compressive projections onto different FMLO channels. (d) When the frequency interval $\delta f_1$ = 50 kHz is complementary to $f_{\rm{mod}}$, e.g., $\delta f_1  = 50~\rm{kHz} $ (blue) and $\delta f_1  = 76~\rm{kHz} $ (orange) with $f_{\rm{mod}}$ = 126 kHz, the resulting compressive measurements exhibit nearly identical frequency structures, differing only in the relative projection weights.}
        \label{fig2}
    \end{figure*}

    \begin{figure*}
        \centering
        \includegraphics[width=1\linewidth]{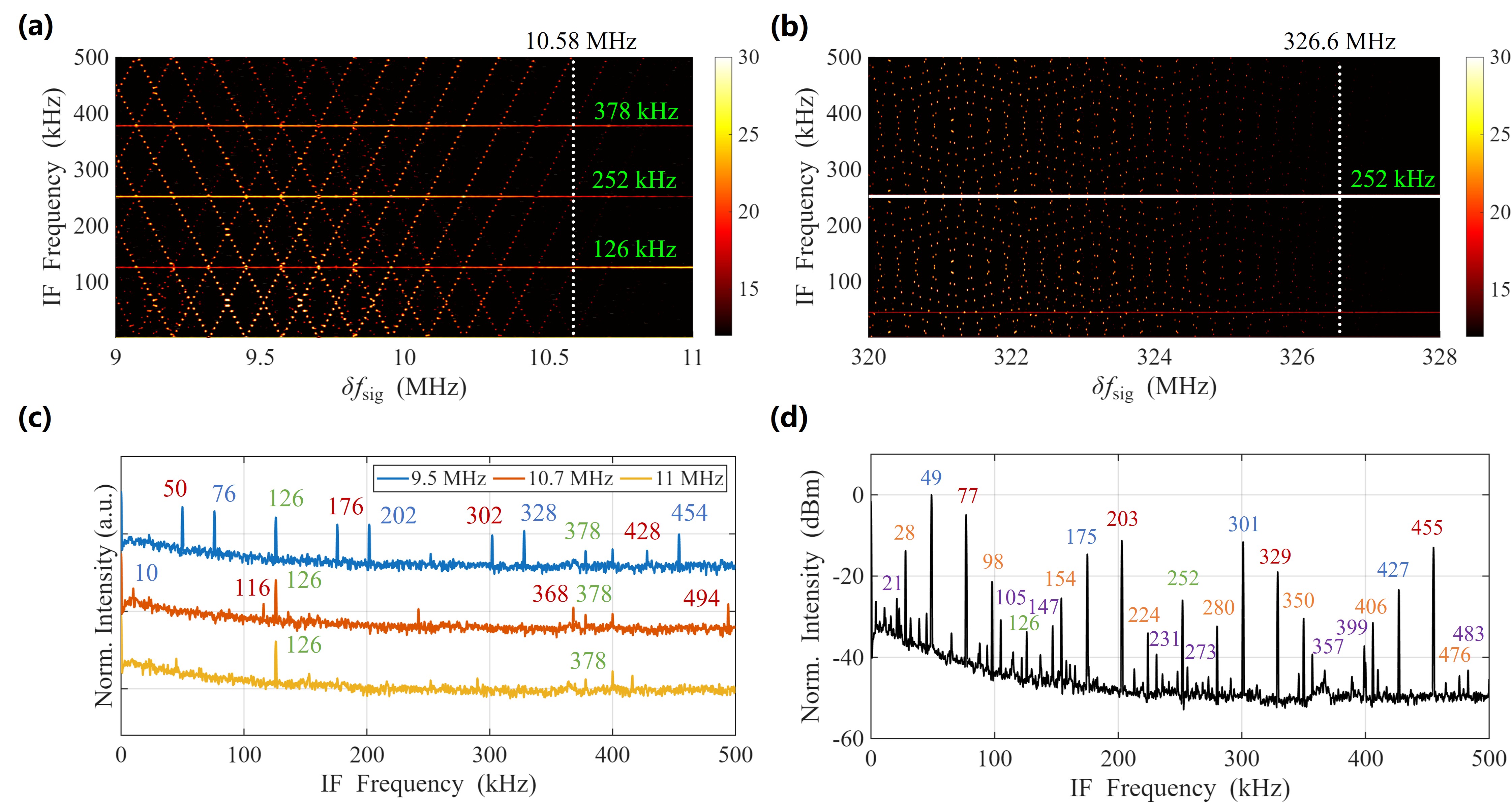}
        \caption{\textbf{Recoverable spectral range of the compressive receiver.} (a) Compressive IF spectra as a function of the signal detuning $\delta f_{\rm{sig}}$ for $f_{\rm{dev}} = 10$~MHz and $f_{\rm{mod}} = 126$~kHz. The white dashed lines indicate the maximum detectable detuning. (b) IF spectra map for an FMLO with $f_{\rm{dev}} = 320$~MHz and $f_{\rm{mod}} = 252$~kHz, demonstrating the extended single-side recoverable spectral range up to $\sim 326.6$~MHz. (c) IF spectra for $\delta f_{\rm{sig}} = 9.5$~MHz (blue), $10.7$~MHz (orange), and $11$~MHz (yellow) with $f_{\rm{dev}} = 10$~MHz, showing the degradation of compressive measurements near the FMLO boundary (d) Higher-order mixing products observed under an increased signal power. Peaks in blue, red, and green correspond to the second-order nonlinear interactions, while the orange and purple ones indicate the third- and the fourth-order interactions. Additional weak mixing products from even higher-order nonlinear processes are also present in the spectrum but are not explicitly labeled.}
        \label{fig3}
    \end{figure*}

    \begin{figure*}
        \centering
        \includegraphics[width=1\linewidth]{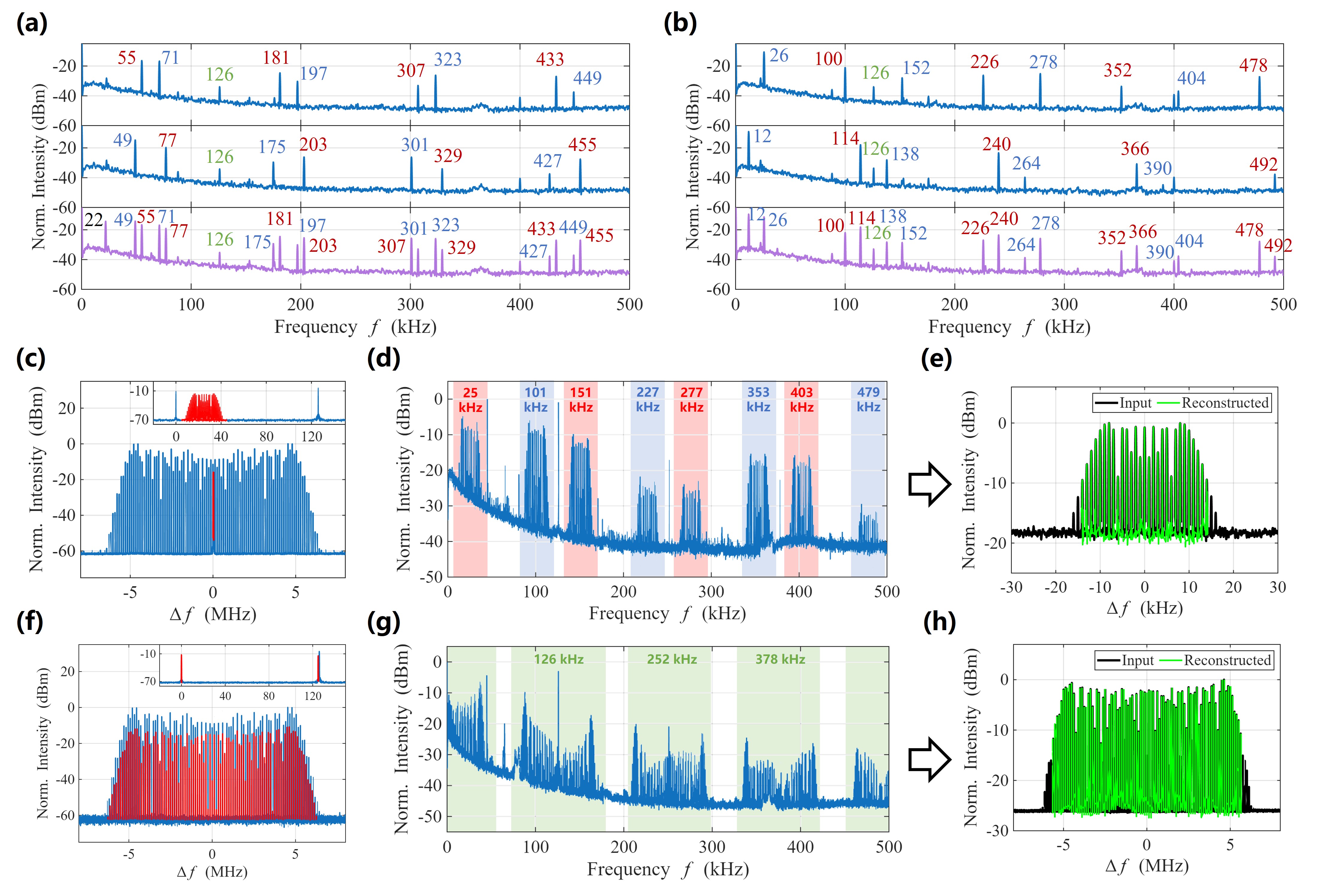}
        \caption{\textbf{Multi-frequency response and broadband FM signal recovery.} (a) Response spectra for two single-frequency signal fields with $\delta f_{\rm{sig}} = 55$~kHz (upper panel) and 77~kHz (middle panel) under an FMLO with $f_{\rm{mod}} = 126$~kHz. The lower panel (purple) shows the simultaneous response, where the peak at $22$~kHz (black label) arises from mixing between the two tones. (b) Same configuration but with $\delta f_{\rm{sig}} = 100$~kHz and $996$~kHz, corresponding to $\delta f_1 = 100$~kHz and $114$~kHz. The simultaneous response remains a linear superposition of the individual responses. (c) Spectral arrangement of a narrowband FM signal (red, frequency deviation 10~kHz, modulation rate 1~kHz, $\delta f_{\rm{sig}} = 25$~kHz) and the FMLO spectrum (blue). The inset shows a magnified view from $-20$~kHz to 150~kHz. (d) Output spectrum under the narrowband FM signal configuration, exhibiting a series of IF signals with an FM-like structure. (e) Reconstructed FM signal spectrum (green) compared with the input spectrum (black). The reconstruction incorporates power compensation on each comb component according to the FMLO spectral envelope. (f) Spectral arrangement of a broadband FM signal (red, frequency deviation 5~MHz, modulation rate 125~kHz, $\delta f_{\rm{sig}} = 0$~kHz) and the FMLO spectrum (blue). The inset shows a magnified view from $-20$~kHz to 150~kHz. (g)  Output spectrum under the broadband FM signal configuration. Each frequency component of the FM signal mixes with all frequency components of the FMLO, producing a superposed response spectrum with features spaced by the difference between the two modulation rates. (h) Reconstructed FM signal spectrum (green) derived from the response features near $252$~kHz in (g), compared with the input spectrum (black). The reconstruction incorporates power compensation based on the FMLO spectral envelope.}
        \label{fig4}
    \end{figure*}

    \begin{figure*}
        \centering
        \includegraphics[width=1\linewidth]{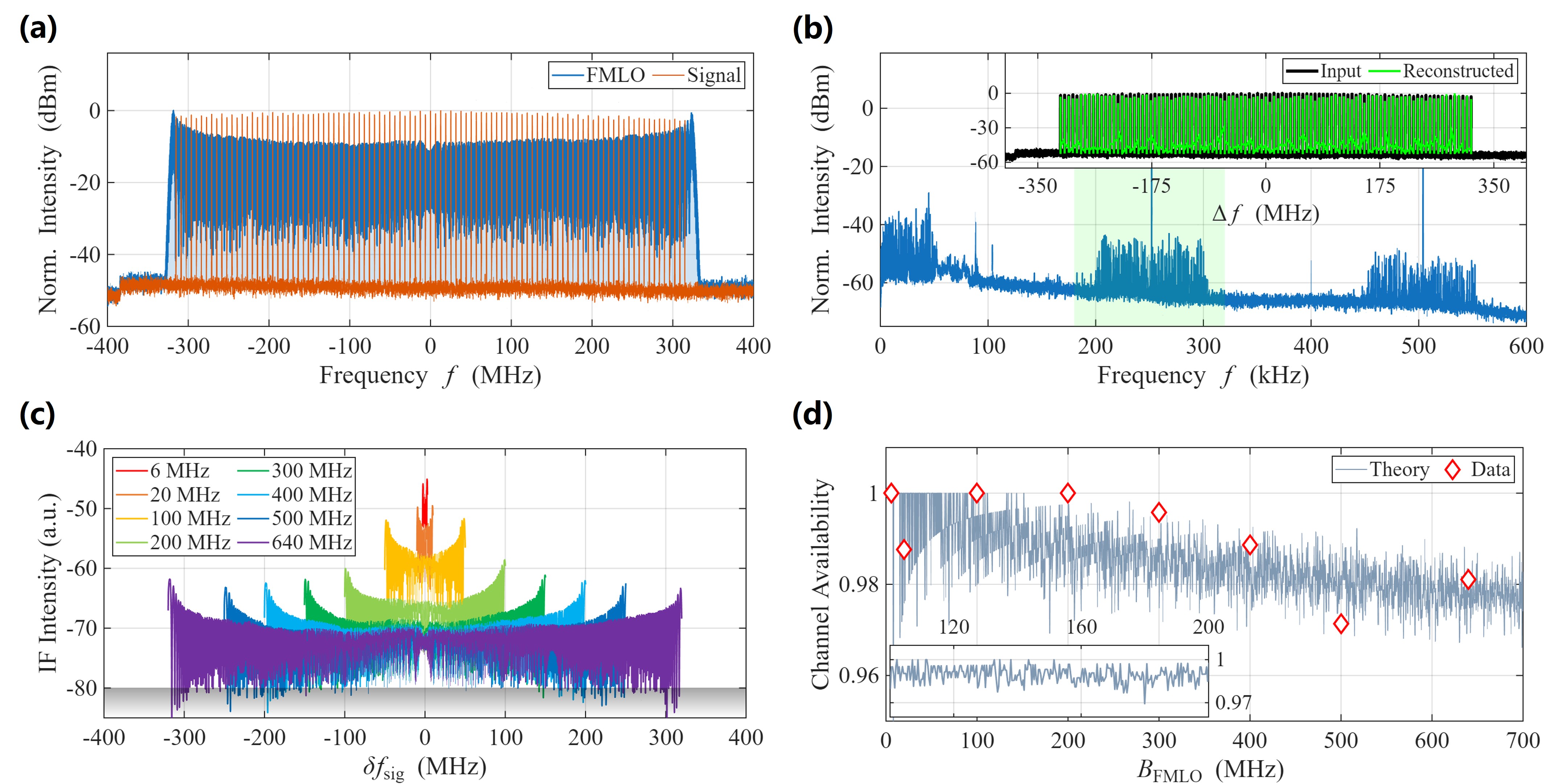}
        \caption{\textbf{Recoverable spectral range exceeding 640 MHz achieved through spectral compression and the associated channel availability.} (a) Spectral arrangement of the FMLO with a frequency deviation of 320 MHz and a modulation rate of 252 kHz (blue) and a 101-tone multi-frequency signal spanning from -320 MHz to 320 MHz with an equal spacing of 6301 kHz (orange). (b) The measured intermediate frequency spectrum of the atomic response (blue). The inset shows the reconstructed signal spectrum (green) obtained from the IF replicas within the 252 kHz region highlighted by the green shading which agrees well with the programmed input spectrum (black). (c) Response amplitude of individual FMLO channels to a single-frequency signal with a fixed power and a constant frequency offset of 50 kHz under varying FMLO frequency deviations. The gray shading indicates the threshold below which the channel response is considered too weak for reliable compressive projection. (d) Channel availability as a function of the FMLO spectral width. The experimental results (orange circles) agree with the theoretical prediction (gray solid line). The theoretical curve exhibits a fine structure originating from the alternating signs and varying magnitudes of different-order Bessel functions of the first kind. The inset provides a magnified view of the 100 MHz to 200 MHz range.}
        \label{fig4a}
    \end{figure*}

    \begin{figure*}
        \centering
        \includegraphics[width=1\linewidth]{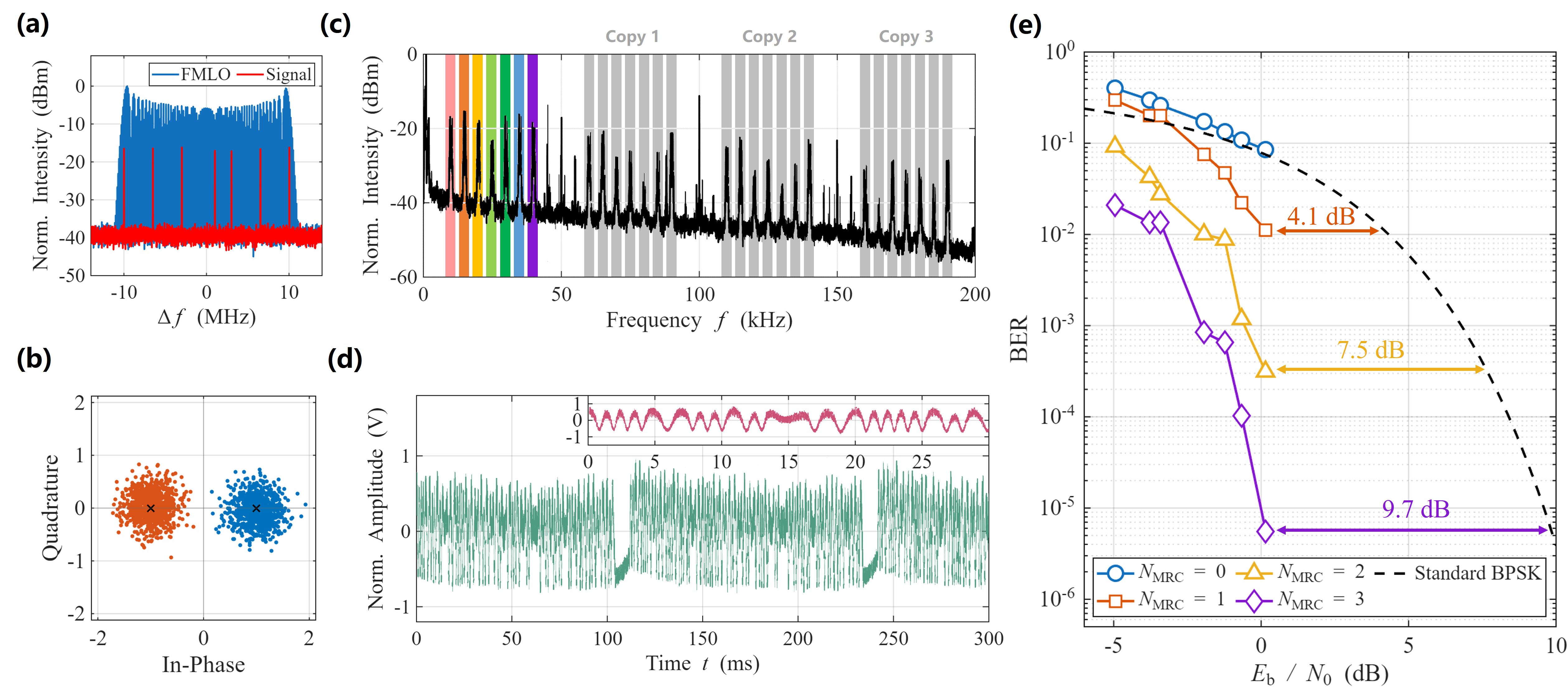}
        \caption{\textbf{Multi-channel communication enabled by the FMLO architecture.} (a) Schematic of the relative spectral arrangement between the FMLO and the multi-frequency signal field. The FMLO has a frequency deviation $f_{\rm{dev}} = 10$~MHz and a modulation rate $f_{\rm{mod}} = 100$~kHz. The multi-frequency signal is a composite waveform carrying 1~kbps BPSK-modulated data on seven carriers, located at frequency offsets of $-10$~MHz~$+$~10~kHz, $-6.5$~MHz~$+$~15~kHz, $-3$~MHz~$+$~20~kHz, $+1$~MHz~$+$~25~kHz, $+3$~MHz~$+$~30~kHz, $+6.5$~MHz~$+$~35~kHz, and $+10$~MHz~$+$~40~kHz relative to the FMLO center frequency. Each carrier constitutes an independent communication channel and is modulated with a distinct pseudo-random bit sequence. (b) Constellation diagram of 1500~bits within a single communication frame. Black crosses mark the two standard bit states at $(\pm 1, 0)$, while orange and blue points represent bits decoded as $-1$ and $+1$, respectively. (c) Frequency-domain spectrum of the composite IF signal obtained from the Rydberg atomic receiver when all seven channels are transmitted simultaneously. The direct mixing products (10~kHz to 40~kHz) corresponding to different channels are highlighted by rainbow-colored shaded regions (red, orange, yellow, green, cyan, blue, and violet). Gray shaded regions mark additional IF replicas that provide MMV of the same transmitted symbols. (d) Time-domain waveform of the composite IF signal with all seven channels active. The inset shows an enlarged view of the IF signal over a 0-30~ms window. (e) Measured BER as a function of the equivalent bit-energy-to-noise-power-density ratio $(E_b/N_0)$, benchmarked against the theoretical BER curve of BPSK over an AWGN channel (black dashed line). When only a single IF response is used for demodulation (blue circles), the BER curve closely follows the standard BPSK reference. In contrast, MRC combining of multiple IF replicas yields a significant BER reduction, demonstrating the gain from coherently combining multiple compressive projections of the same signal. The performance improves monotonically with the number of replicas ($N_{\rm{MRC}}$).}
        \label{fig5}
    \end{figure*}

\section*{Result}

    \textbf{Compressive Spectral Multiplexing Model}
    Conventional heterodyne detection employs a single-frequency LO that mixes with the incident signal, generating an IF response bounded by the atomic instantaneous bandwidth. In contrast, the FMLO utilizes multiple equally spaced, phase-locked spectral tones, each serving as an independent mixing channel subject to the same bandwidth constraint. The resulting multi-heterodyne process collectively maps the broadband spectral information of the input signal into multiple narrowband IF outputs, realizing a physical compressive measurement. Accordingly, the FMLO channel ensemble defines a sensing matrix $\bf{A}$ and the multi-heterodyne mixing effectively performs a dimensionality reduction $\bf{y} = \bf{A x} + \bf{n}$, where $\bf{x}$ is the sparse input spectrum, $\bf{y}$ denotes the measured IF spectrum, and $\bf{n}$ represents additive noise. Since the number of IF outputs is much smaller than the spectral degrees of freedom, recovering $\mathbf{x}$ from $\mathbf{y}$ is an underdetermined inverse problem rendered tractable by the sparsity of $\mathbf{x}$.
    
    We first characterize the structure of the multiplexed channels. The FMLO is a sinusoidal FM signal with constant amplitude, with its time-domain and frequency-domain forms given by
    \begin{align}
        E_{\rm{LO}}\left(t\right) = E_c \cos\left[2\pi \it{f_c}t + \beta \rm{sin} \left(2\pi \it{f_{\rm{mod}}}t\right)\right]
        \label{eq1}
    \end{align}
    \begin{align}
        E_{\rm{LO}}\left(f\right) = \frac{E_c}{2} \sum_{m=-\infty }^{+\infty} J_m \left( \beta\right) \delta\left[f-\left(f_c + mf_{\rm{mod}}\right) \right]
        \label{eq2}
    \end{align}
    where $\beta = f_{\rm{dev}}/f_{\rm{mod}}$ is the modulation index, $f_{\rm{dev}}$ is the frequency deviation, $f_{\rm{mod}}$ is the modulation rate, $f_c$ is the center frequency, and $J_m$ is the $m$-th order Bessel function of the first kind. Schematic representations of the FMLO in the time and frequency domains are illustrated in Fig. \ref{fig1} (a). The frequency, amplitude, and phase of the individual components follow
    \begin{equation}
        \begin{cases}
            f_m = f_c + mf_{\rm{mod}}\\
            E_m = \frac{1}{2}E_c J_m\left ( \beta \right )  \\
            \varphi_m = \arg\left [ \it{J_m} \left ( \beta \right ) \right ],
        \end{cases}
        \label{eq3}
    \end{equation}  
    where $m$ indexes the frequency components and $\arg[\cdot]$ denotes the phase angle. The resulting channel ensemble spans a bandwidth $B_{\rm{FMLO}} \approx 2f_{\rm{dev}}$ with approximately $N = 2\beta + 1$ effective components according to the Carson formula, all intrinsically phase-locked, in contrast to architectures based on multiple independent LO sources.

    In our experiment, Rydberg atoms are prepared via two-photon excitation using a 780 nm probe laser and a 480 nm coupling laser. The FMLO and the external signal are combined through a resistive power divider and coupled into the vapor cell via a waveguide, as illustrated in Fig.~\ref{fig1} (b). The global spectral mapping enabled by the FMLO operates as follows. As shown in Fig.~\ref{fig1} (c1), a broadband multi-tone signal overlaps with the FMLO spectrum and undergoes multi-heterodyne mixing in Rydberg atoms. Owing to the finite response bandwidth of each atomic channel, different spectral components are selectively down-converted by different FMLO tones. The aggregated response compresses the broadband input into a set of narrowband intermediate-frequency (IF) components, as illustrated in Fig.~\ref{fig1} (c2), forming multiple spectral projections of the input signal. To elucidate the local mixing mechanism underlying this compression, we consider a single-frequency input signal at $f_{\rm{sig}}$, with frequency offset $\delta f_{\rm{sig}} = f_{\rm{sig}} - f_c$ relative to the center frequency of the FMLO. The IF output can be expressed as
    \begin{equation}
        P_{\rm{IF}}(t) \propto \sum_{m=-\infty}^{+\infty} E_m E_{\rm{sig}} \cos\left[2\pi (f_m - f_{\rm{sig}}) t + \varphi_m\right],
        \label{eq4}
    \end{equation}
    yielding IF components at
    \begin{equation}
        f_{\rm{IF}}^{(m)} = |\delta f_{\rm{sig}} - m f_{\rm{mod}}|.
        \label{eq5}
    \end{equation}
    $\delta f_1$ and $\delta f_2$ denote the frequency separations from the signal to its two nearest channels, satisfying $\delta f_1 + \delta f_2 = f_{\rm{mod}}$. The IF spectrum then consists of three families, namely $\delta f_1 + n f_{\rm{mod}}$, $\delta f_2 + n f_{\rm{mod}}$, and $n f_{\rm{mod}}$ for $n \in \mathbb{N}$, as illustrated in Fig.~\ref{fig1} (c3, c4). A single frequency input signal therefore generates multiple IF replicas across the channel ensemble. Owing to the mirror symmetry of the Bessel envelope about $f_c$, signals at complementary detunings produce identical frequency distributions in the IF output, differing only in the relative intensities governed by $J_m(\beta)$. Although each channel is individually bandwidth-limited, different spectral components of a broadband input are distributed across distinct channels. When the resulting IF components remain spectrally separable, the input spectrum can be reconstructed from the compressive measurements in principle.

    Moreover, the multiplexing architecture decouples the sensing bandwidth from the atomic instantaneous bandwidth. Each channel operates within a narrow bandwidth centered at $f_m$, while the channel ensemble spans the FMLO spectral envelope. The recoverable bandwidth is therefore $B_{\rm{recover}} \approx B_{\rm{FMLO}} \approx 2f_{\rm{dev}}$, determined solely by the FMLO modulation parameters and bounded above by approximately $N$ times the intrinsic atomic bandwidth when $f_{\rm{mod}}$ is set to twice that bandwidth. This spectral coverage is realized by the channel ensemble functioning as a sensing matrix that compresses broadband spectral information into multiple narrowband IF outputs, enabling a narrowband atomic receiver to access a much broader spectral range. A single component of the input signal is mapped to multiple IF replicas occupying distinct frequency bands, forming a multiple measurement vector (MMV) acquisition with statistically uncorrelated noise across replicas. This redundancy enables coherent combining to improve the detection performance. To validate this benefit for communication signals, we consider a transmitted symbol $S$ observed through $M$ IF replicas, where the value of $M$ depends on $\beta$ and the position of the signal center frequency relative to the FMLO spectrum. The received signal from the $m$-th replica can be modeled as 
    \begin{equation}
        r_m = h_m S + n_m,
        \label{eq6}
    \end{equation}
    where $h_m$ is the channel gain scaling with $|J_m(\beta)|$, and $n_m \sim \mathcal{N}(0, \sigma_m^2)$ is the mutually uncorrelated noise across replicas. The SNR of each replica is $\gamma_m = h_m^2 / \sigma_m^2$. With maximal-ratio combining (MRC) technique, the $M$ replicas are weighted by their channel gains and coherently summed, yielding the combined symbol estimate $\hat{S} = \sum_m h_m r_m / \sum_m h_m^2$. The effective SNR after MRC is the sum of the individual replica SNRs,
    \begin{equation}
        \gamma_{\rm{MRC}} = \sum_{m=1}^{M} \gamma_m.
        \label{eq7}
    \end{equation}
    The increase in $\gamma_{\rm{MRC}}$ directly improves the detection robustness, with each additional replica contributing its own SNR to the combined output.

    \textbf{Single-Frequency Response}
    The structured spectral multiplexing model predicts that the FMLO channel ensemble defines a sensing matrix $\mathbf{A}$ and that the multi-heterodyne mixing performs a compressive mapping $\mathbf{y} = \mathbf{A}\mathbf{x} + \mathbf{n}$. We experimentally verify this model in the 1-sparse limit (single-frequency input). In this regime, the input spectrum $\mathbf{x}$ contains a single nonzero component, while $\mathbf{y}$ consists of multiple IF replicas corresponding to projections onto different rows of $\mathbf{A}$, directly revealing the sensing matrix structure.

    The FMLO follows sinusoidal modulation, yielding a Bessel-distributed spectral envelope (Eq.~\ref{eq2}), with measured spectrum shown in Fig.~\ref{fig2}(a). Our Rydberg system features a narrow intrinsic instantaneous bandwidth of approximately 126 kHz for off-resonant frequencies, as characterized in Fig.~\ref{fig2} (b). 
    
    When only the FMLO is coupled into the atomic vapor cell, all FMLO frequency components undergo mutual mixing through the AC Stark effect, producing IF signals at the integer multiples of $f_{\rm{mod}}$. The resulting IF spectrum exhibits components at $f_{\rm{mod}}$, $2f_{\rm{mod}}$, $3f_{\rm{mod}}$, and higher-order harmonics, consistent with the multi-channel mixing mechanism described by the compressive sensing model. Once an additional signal field is applied, the atoms mix it with all FMLO frequency components and generate the cross terms in Eq.~(\ref{eq4}). For $f_{\rm{dev}} = 3$~MHz and $f_{\rm{mod}} = 126$~kHz, a signal with $\delta f_{\rm{sig}} = 70$~kHz produces the IF spectrum shown in Fig.~\ref{fig2}(c). The output consists of three frequency families, $\delta f_1 + m f_{\rm{mod}}$, $\delta f_2 + m f_{\rm{mod}}$, and $m f_{\rm{mod}}$, with $\delta f_1 + \delta f_2 = f_{\rm{mod}}$. These components take the form $|k \delta f_{\rm{sig}} + m f_{\rm{mod}}|$ with $k = 0, \pm1$, representing distinct projections of the same 1-sparse signal. 
    Given the symmetry of the FMLO spectrum, signals at detunings complementary to $f_{\rm{mod}}$ produce IF spectra with identical frequency distributions but different amplitudes. This is illustrated in Fig.~\ref{fig2} (d) for $\delta f_1 = 50$ kHz (blue) and $\delta f_1 = 76$ kHz (orange), where the intensity relationship at 428 kHz and 454 kHz is reversed between the two cases. This difference arises from the nonuniform intensity distribution of the FMLO spectrum in combination with its inherent symmetry. Further details can be found in the Supplementary Information.

    Compared to single frequency LO heterodyne detection, where the response vanishes once $\delta f_{\rm{sig}}$ exceeds the instantaneous bandwidth, the response bandwidth of the FMLO-based receiver is limited by the spectrum span of the sensing matrix. As shown in Fig.~\ref{fig3} (a)–(c), the IF output persist across the FMLO spectral envelope and gradually diminishes near its boundary. This coverage can be expanded by increasing $f_{\rm{dev}}$, consistent with $B_{\rm{recover}} \approx 2 f_{\rm{dev}}$. In our experiment, the maximum $f_{\rm{dev}}$ obtained is 320 MHz, limited by the microwave source. With better instruments, $f_{\rm{dev}}$ can be increased to several gigahertz. For brevity and clarity, Figs.~\ref{fig3} (a) and (b) illustrate only the system response near the spectral edge of the FMLO spectrum, with the complete IF response for the 320~MHz FMLO configuration shown in the Supplementary Information. Figure~\ref{fig3} (c) compares the compressive IF spectra for signals with the center frequencies positioned within, near the edge of, and beyond the FMLO spectral coverage. As $\delta f_{\rm{sig}}$ exceeds the sensing matrix boundary, the projection amplitudes decrease with stronger attenuation at lower orders, therefore defining the upper limit of the recoverable spectral range. Additionally, under strong signal fields, the system exhibits higher-order mixing products where $k$ in $|k\delta f_{\rm{sig}}+mf_{\rm{mod}}|$ is no longer limited to 0 or $\pm1$ but can take absolute values of 2 or even larger, as shown by the peaks marked in orange and purple in Fig.~\ref{fig3} (d). These higher-order products correspond to nonlinear extensions of the sensing matrix and can be exploited to enhance the frequency measurement precision (see Methods).

    \textbf{Multi-Frequency Response and Discrimination}
    The single-frequency measurements verify the compressive sensing model in the 1-sparse limit. For a multi-tone input where frequency components are weak enough to render their intermodulation negligible, the compressive measurement is linear, i.e., the IF spectrum is the sum of the projections of each individual tone. We verify this property using a dual-tone signal, with experimental results shown in Fig.~\ref{fig4} (a). With $f_{\rm{mod}}$ = 126~kHz and $f_{\rm{dev}}$ = 1~MHz, the response spectra for individual tones at \(\delta f_{\rm{sig}} = 55\) kHz and \(\delta f_{\rm{sig}} = 77\) kHz are shown in the upper and middle panels, respectively. When both frequency components are present in the input signal simultaneously, the frequency components of the resulting response spectrum (lower panel) precisely match the superposition of the individual responses. The peak at $22$~kHz (black labels) originates from intermodulation between the two signal fields. Figure~\ref{fig4} (b) confirms the same linearity for tones at $\delta f_{\rm{sig}} = 100$~kHz and 996~kHz, far from the complementary configuration in (a). This linear superposition property holds across the entire spectral range covered by the FMLO, establishing that the compressive measurement model $\bf{y = A x + n}$ is valid for spectrally sparse multi-tone inputs. The linearity of the compressive measurement enables frequency estimation from the IF spectrum. In the weak-signal regime, the lowest-frequency component $f_{\rm{min}}$ satisfies $f_{\rm{min}} \equiv \pm \delta f_{\rm{sig}} \pmod{f_{\rm{mod}}}$, which determines the offset of $\delta f_{\rm{sig}}$ from the nearest integer multiple of $f_{\rm{mod}}$, up to a sign ambiguity. A single measurement therefore yields a discrete set of candidate frequencies. This ambiguity can be resolved by performing a second measurement with a different modulation rate $f_{\rm{mod}}'$, and the true $\delta f_{\rm{sig}}$ can then be recovered as the unique value satisfying both congruence relations over the unambiguous range determined by the least common multiple of the two modulation rates. By further casting the estimation as a robust optimization over the measured $f_{\rm{min}}$ values, this approach achieves broadband frequency estimation with high precision even in the presence of measurement uncertainties. A detailed description of the estimation algorithm is provided in the Methods.
    
    For input signals with multiple frequency components, Figs.~\ref{fig4} (c)-(e) demonstrate signal recovery from the compressive IF measurements. We consider an FM signal with a discrete, equally spaced spectrum. In a conventional Rydberg receiver with a single-frequency LO, such an FM signal can only be received if its entire spectrum falls within the atomic instantaneous bandwidth. Using the FMLO-based compressive receiver, however, the signal field mixes with all FMLO frequency components, producing a series of IF outputs with similar spectral distributions, as shown in Fig.~\ref{fig4} (d). While infinite atomic bandwidth yields a complete compressed replica in each output, finite bandwidth reduces per-channel information, yet the original spectrum remains recoverable through the combination of multi-channel responses. Figure~\ref{fig4} (e) shows the recovered FM signal spectrum from the IF component centered near 151~kHz [red shaded region in Fig.~\ref{fig4} (d)], compared with the original input spectrum. The recovery incorporates power compensation according to the known sensing matrix envelope. A practical constraint arises when the spectral coverage of the signal exceeds the frequency difference between its center frequency and the nearest FMLO comb line, causing the IF outputs to alias. In such cases, increasing $f_{\rm{mod}}$ widens the alias-free interval, but at the cost of compromised measurement quality. Once $f_{\rm{mod}}$ exceeds twice the intrinsic atomic bandwidth, the response of individual channels becomes significantly weaker. This trade-off reflects a fundamental limit of the compressive architecture. The intrinsic atomic bandwidth determines the maximum rate at which independent compressive projections can be acquired, which in turn bounds the maximum alias-free signal bandwidth that can be recovered.
    
    Despite this constraint, for multi‑frequency signals with discrete, sparse spectral occupancy across a broad frequency range, the FMLO-enabled compressive receiver can recover signals well beyond the intrinsic atomic instantaneous bandwidth, provided the IF spectrum remains alias-free. We demonstrate this using a broadband FM signal with a bandwidth of 10~MHz and a modulation rate of 125~kHz, while applying an FMLO with a frequency deviation of 10~MHz and a modulation rate of 126~kHz, as shown in Fig.~\ref{fig4} (f). The slight difference between the two modulation rates creates a dual-comb-like multi-heterodyne mapping that compresses the broadband FM spectrum into a set of narrowband IF signals spaced by 1~kHz, as indicated by the green shaded region in Fig.~\ref{fig4} (g). Selecting the IF output with the highest SNR, we perform power compensation using the known sensing matrix envelope to recover the original FM spectrum. The recovered spectrum (green) is compared against the input spectrum (black) in Fig.~\ref{fig4} (h). To further validate the scalability of this approach, we probe the full span of the FMLO spectral envelope with a densely spaced multi-tone signal. As shown in Fig. \ref{fig4a} (a), a 101-tone signal with a uniform spacing of 6301 kHz is distributed across a 640 MHz range, matching the spectral extent of an FMLO with a frequency deviation of 320 MHz. This specific spacing produces IF components separated by 1 kHz, ensuring spectral resolvability within the narrowband atomic response. Following a similar compressive multi-heterodyne process, the resulting IF spectrum is shown in Fig. \ref{fig4a} (b). The input signal spectrum is faithfully reconstructed after applying power compensation to different channels according to the envelope of the sensing matrix, confirming a recoverable spectral range exceeding 640 MHz within a single acquisition frame.

    The deterministic structure of the sensing matrix also governs the effective number of channels available for reliable compressive projection. Since the amplitude of each FMLO tone follows the Bessel function distribution described by Eq.~\ref{eq3}, channels with weaker LO amplitudes exhibit correspondingly weaker responses. We therefore measured the response of individual FMLO channels to a calibration signal with a fixed power and a constant frequency offset of 50 kHz under varying FMLO frequency deviations, as shown in Fig. \ref{fig4a}(c). As the FMLO spectral width increases, the overall response amplitude gradually declines because the fixed total LO power is distributed among a growing number of frequency components. Channels whose response falls below a threshold are considered too weak for reliable projection, and the channel availability is defined as the fraction of channels exceeding this threshold. The experimental results shown in Fig. \ref{fig4a}(d) agree with the theoretical prediction derived from the Bessel envelope of the FMLO spectrum. The theoretical curve (gray solid line) exhibits a fine structure that reflects the alternating signs and varying magnitudes of different-order Bessel functions of the first kind evaluated at the same modulation index. We note that this structure depends on the modulation parameters, and a different choice of $f_{\rm{mod}}$ would yield a distinct channel envelope.

    \textbf{Multi-Channel Communication Demonstration}
    The compressive spectral multiplexing framework can be naturally applied for multi-channel communication, where the sensing matrix maps the narrowband signals from different communication channels into distinct IF bands. This mapping arises from the multi-heterodyne mixing with the FMLO and enables multi-channel communication with a single atomic receiver. We demonstrate this capability using seven concurrent BPSK channels, as shown in Fig.~\ref{fig5}. Each channel possesses one carrier and carries a 1~kbps BPSK pseudo-random bit sequence. Each carrier is assigned a unique frequency offset relative to its nearest FMLO comb line, such that the down‑converted IF signals occupy distinct, non‑overlapping frequency bands and can be readily separated by band-pass filtering. The FMLO parameters are $f_{\rm{dev}} = 10$ MHz and $f_{\rm{mod}} = 100$ kHz, with the relative spectral arrangement between the communication carriers and the FMLO illustrated in Fig.~\ref{fig5} (a). Figure~\ref{fig5} (b) shows the demodulated constellation with $10^4$ symbols, where the dispersion around $(\pm 1, 0)$ reflects the channel SNR.

    The frequency- and time-domain waveform of the composite IF signal are shown in Figs.~\ref{fig5} (c) and \ref{fig5} (d), respectively. In the frequency domain, the seven rainbow-colored regions correspond to direct IF outputs of the individual channels. Consistent with the multi-heterodyne mixing mechanism characterized in Fig.~\ref{fig4}, three additional replicas (labeled Copy~1 to Copy~3, gray shaded regions) appear between 50~kHz and 200~kHz. These replicas constitute a MMV realization, where the same transmitted symbol is observed across multiple frequency-separated projections. As these replicas occupy distinct frequency bands, their noise contributions are uncorrelated, enabling coherent combining to improve detection reliability. To quantify this MMV gain, Fig.~\ref{fig5}(e) presents the measured bit error rate (BER) versus the equivalent bit-energy-to-noise-power-density ratio $(E_b/N_0)$, benchmarked against the theoretical BER for BPSK over an AWGN channel (black dashed line), with measurements limited to $E_b/N_0 \lesssim 0$ dB due to the large data volume required for reliable estimation at lower BER. When only a single IF response is used for demodulation, without invoking any replica, the resulting BER curve (blue circles) closely follows the standard BPSK limit. In contrast, applying maximal-ratio combining (MRC) across multiple replicas significantly improves the performance. The BER decreases monotonically with the number of combined replicas $N_{\rm{MRC}}$. The system with $N_{\rm{MRC}} = 3$ achieves the same BER at $E_b/N_0 \approx 0$ dB, comparable with single-channel BPSK at $E_b/N_0 \approx 10$ dB, corresponding to an effective $\sim$10 dB reduction in the required $E_b/N_0$ for the same BER. This improvement arises from the coherent combining of independent noisy projections, where each replica contributes additively to the overall SNR as described by Eq.~(\ref{eq7}). These results demonstrate that the compressive multiplexing architecture provides not only access to an expanded spectrum range but also a tangible diversity gain that enhances communication reliability even under severely degraded channel conditions.
    
\section*{Discussions and Summary}

    Compressive sensing provides a framework for acquiring high-dimensional signals through a reduced number of measurements by exploiting the spectral sparsity, typically relying on engineered random projections followed by numerical reconstruction. In contrast, our FMLO-based receiver realizes a structured form of spectral compression in which the broadband input spectrum is physically mapped onto multiple narrowband outputs through multi-heterodyne mixing. The resulting measurement can be formulated as a linear model with a deterministic sensing matrix defined by the FMLO spectrum, in which compression and acquisition take place concurrently within the atomic medium, eliminating the need of engineered randomness or iterative recovery. Within this framework, the FMLO generates a set of phase-locked frequency components with fixed amplitude and phase relations that collectively define the sensing matrix. In contrast to microwave frequency comb approaches relying on independently controlled comb lines, the FMLO enables a compact, low-cost, and readily accessible implementation, with relative channel weights governed by the Bessel envelope. Although this inherent nonuniformity leads to unequal channel responses and limited independent control over individual frequency components, it directly encodes the sensing matrix structure and can be compensated during signal reconstruction (as discussed in the Supplemental Material), while substantially reducing system complexity and calibration overhead.

    The achievable spectral coverage is determined by the FMLO bandwidth, and the present implementation achieves a total recoverable bandwidth exceeding 640 MHz. We note that extending the FMLO bandwidth can further increase the recoverable range albeit at the cost of increased power requirements, since a larger number of frequency components results in reduced power per channel and consequently weaker individual responses. This trade-off between spectrum coverage and single-channel sensitivity defines a practical limit on the usable FMLO bandwidth. However, combining the present scheme with approaches that increase the intrinsic atomic instantaneous bandwidth allows larger spacing between the FMLO frequency components, thereby reducing the number of required components for a given coverage. This alleviates the power distribution constraint and improves the efficiency of broadband sensing, indicating that the atomic bandwidth primarily governs the efficiency of spectral compression rather than the recoverable spectrum range itself.

    Meanwhile, the multi-channel nature of the FMLO establishes a regime in which multiple projections of the same signal are acquired simultaneously. While the 101-tone demonstration was restricted to $\pm 320$ MHz by the bandwidth of our microwave source, independent single-tone channel-response measurements (Fig.~\ref{fig3}) indicate that the intrinsic recoverable limit of this FMLO configuration extends to approximately $\pm 326$ MHz. Moreover, as evidenced by the broadband FM recovery in Fig.~\ref{fig4}, a larger number of tones can in principle be accommodated, provided that the down-converted IF components remain free of mutual aliasing. This multiple measurement structure furnishes intrinsic redundancy that can be exploited to enhance the detection SNR, and the observed improvement in communication reliability arises from MRC of independent projections.

    In summary, we have proposed and experimentally demonstrated a compressive broadband spectrum sensing architecture for Rydberg atomic receivers based on an FMLO. By replacing the conventional single-tone LO with an FMLO, the intrinsic multi-heterodyne mixing process is harnessed for spectral compression, enabling a spectrum recovery range exceeding 640 MHz, more than three orders of magnitude broader than the intrinsic atomic instantaneous bandwidth. We have systematically characterized the underlying multi-heterodyne mixing mechanism and verified the compressive measurement model in both the 1-sparse limit and multi-tone sparse regimes. In the 1-sparse limit, the FMLO architecture naturally generates multiple IF replicas for single-frequency inputs, while in the multi-tone sparse regime it compresses broadband multi-tone spectra into several narrowband IF responses, each constituting a complete projection of the original signal. Therefore, the deterministic sensing matrix defined by the FMLO provides both a broader spectral access and an intrinsic measurement redundancy, enhancing the receiver performance and robustness for broadband applications. For instance, a 10 dB diversity gain is obtained for a multi-channel communication scenario. Moreover, this approach requires only a single microwave waveform and no auxiliary fields or broadband electronics, offering a scalable pathway toward compact quantum receivers for broadband sensing and next-generation wireless communication systems.

\section*{Methods}
    
    \textbf{Experimental setup}
    $^{85}$Rb atoms in a room-temperature vapor cell are excited to the Rydberg state via a two-photon EIT scheme using a probe laser (780 nm, beam waist 320 \(\mu\)m) and a coupling laser (480 nm, beam waist 560 \(\mu\)m), addressing the $5\mathrm{S}_{1/2} \rightarrow 5\mathrm{P}_{3/2} \rightarrow 58\mathrm{D}_{5/2}$ transitions. The energy-level configuration is illustrated in Fig.~\ref{fig1} (a). The corresponding Rabi frequencies for the probe and coupling laser are $\Omega_p = 2\pi \times 18.5$ MHz and $\Omega_c = 2\pi \times 7.7$ MHz, respectively. The overall experimental configuration is shown in Fig.~\ref{fig1} (b). A pair of counter-propagating probe (Toptica DL Pro 780) and coupling (Toptica TA-SHG Pro 480) beams passes through the vapor cell. The FMLO and the signal field are combined via a resistive power divider and coupled into the cell through a waveguide, where they interact with the atoms. Under the combined FMLO and signal field, the Rydberg atoms experience an AC Stark shift, with the energy shift $\Delta E$ depending on the square of the total incident electric field amplitude. The resulting IF response is encoded in the probe transmission and detected by a photodetector (Newport 2001-FS-M), followed by data acquisition. For signal generation, an analog signal generator (Rohde $\&$ Schwarz SWA100B) and a vector signal generator (KSW VSG02) are used to produce the FMLO and test signals, respectively. In the multi-channel BPSK communication experiment, a software-defined radio platform (Ettus Research USRP X410) is employed to generate the modulated RF signals.

    \textbf{Multi-Frequency Mixing}
    In the weak-field regime, the interaction between the FMLO and the signal field through the Rydberg atoms can be described by the second-order AC Stark effect, with the Stark shift given by
    \begin{equation}
        \Delta E = \frac{1}{2} \alpha \left| \tilde{E}_{\rm{LO}} + \tilde{E}_{\rm{sig}} \right|^2,
        \label{AC_Stark}
    \end{equation}
    where $\tilde{E}_{\rm{LO}}(t) = \sum_{m} E_m \rm{exp}\left[\rm{i} 2\pi (\it{f_c \rm{+} \it{m f_{\rm{mod}} }\rm{)}\it{t} } \right]$ and $\tilde{E}_{\rm{sig}}(t) = E_{\rm{sig}} \rm{exp}\left[\rm{i} 2\pi  \it{f_{\rm{sig}} t}\right]$ denote the complex envelope of the FMLO and the signal field, respectively. Considering the atomic relaxation time and the response bandwidth of the photodetector, the mixing product between the FMLO and the signal field can be described by the AC cross-term in Eq.~(\ref{AC_Stark}): 
    \begin{equation}
        P_{\rm{IF}}(t) \propto \tilde{E}^*_{\rm{LO}}(t) \tilde{E}_{\rm{sig}}(t) + \tilde{E}_{\rm{LO}}(t) \tilde{E}^*_{\rm{sig}}(t).
    \end{equation}
    Substituting the series form of the FM signal into the cross-term and expanding the resulting expression, we obtain
    \begin{equation}
        P_{\rm{IF}}(t) \propto \sum_{m=-\infty}^{+\infty} E_m E_{\rm{sig}} \cos\left[2\pi (f_m - f_{\rm{sig}})t + \varphi_m\right].
    \end{equation}
    Consequently, the frequency components observable on a spectrum analyzer can be written as
    \begin{equation}
        f_{\rm{SA}}^{(1)}(m) = |\delta f_{\rm{sig}} - m f_{\rm{mod}}| = |f_{\rm{sig}} - f_c - m f_{\rm{mod}}|.
        \label{eq:first_order}
    \end{equation}
    However, as the power of the signal field increases, higher-order nonlinearities of the AC Stark effect become significant and the atomic polarizability $\alpha$ can no longer be treated as a constant. In this regime, the atomic response should be expanded as a power series in the squared magnitude of the total complex envelope:
    \begin{equation}
        P_{\rm{IF}}(t) \propto \sum_{k=1}^{\infty} \gamma_k \left| \tilde{E}_{\rm{LO}}(t) + \tilde{E}_{\rm{sig}}(t) \right|^{2k}.
        \label{eq:taylor_envelope}
    \end{equation}
    Expanding the $k$-th order term yields various cross products. The dominant contribution responsible for the observed higher-order mixing products arises from the term proportional to $\tilde{E}_{\rm{LO}}^k (\tilde{E}_{\rm{sig}}^*)^k$ and its complex conjugate. Substituting the complex envelope of the FMLO, we have
    \begin{equation}
        \begin{aligned}
            \tilde{E}_{\rm{LO}}^k(t) &= \left( \sum_{m=-\infty}^{\infty} E_m e^{\rm{i} 2\pi \it{m} \it{f}_{\rm{mod}} t} \right)^k 
            \\
            &= \sum_{m=-\infty}^{\infty} A_m^{(k)} e^{\rm{i} 2\pi \it{m} \it{f}_{\rm{mod}} t},
        \end{aligned}
    \end{equation}
    where $A_m^{(k)}$ denotes the effective amplitude of the $m$-th harmonic generated by the $k$-th power of the FMLO comb. Multiplying this by $(\tilde{E}_{\rm{sig}}^*)^k = E_{\rm{sig}}^k e^{-\rm{i} 2\pi \it{k} \delta \it{f_{\rm{sig}} t}}$ and taking the real part, the $k$-th order mixing products yield frequencies at
    \begin{equation}
        \begin{aligned}
            f_{\rm{SA}}^{(k)}(m) &= |k \delta f_{\rm{sig}} - m f_{\rm{mod}}| \\
            &= |k(f_{\rm{sig}} - f_c) - m f_{\rm{mod}}|,
        \end{aligned}
        \label{eq:general_order}
    \end{equation}
    where $m \in \mathbb{Z}$ is an integer indexing the comb lines generated by the nonlinear interaction. This result reveals that each nonlinear effect with order $k$ produces a distinct frequency comb with line spacing $f_{\rm{mod}}$, centered at $k(f_{\rm{sig}} - f_c)$. For $k=1$, Eq.~(\ref{eq:general_order}) reduces to the linear multi-heterodyne spectrum given in Eq.~(\ref{eq:first_order}). For $k=2$, the second-order comb yields frequencies at $|2(f_{\rm{sig}} - f_c) \pm f_{\rm{mod}}|$ and their harmonics, which manifest in the measured spectra as $|\delta f_1 - \delta f_2|$ and $f_{\rm{mod}} - |\delta f_1 - \delta f_2|$, where $\delta f_1$ and $\delta f_2$ denote the frequency separations between $f_{\rm{sig}}$ and its two nearest FMLO comb lines. The higher the power of the signal field, the larger the maximum value of $k$ that should be considered. For the case shown in Fig.~\ref{fig3} (d), terms up to $k = 3$ are sufficient to account for the more abundant mixing products observed in the output signal spectrum.

    \textbf{Multi-tone Signal Spectrum Estimation}
    For a Rydberg microwave receiver operating with a single-frequency LO, the frequency sensing range is limited by its instantaneous bandwidth. To estimate the frequency of a broadband signal, the LO needs to be continuously tuned. In contrast, our FMLO scheme enables real-time broadband spectrum estimation, with its operating principle analogous to the frequency-sensing schemes based on microwave frequency combs \cite{zhang2022Rydberg, chenInstantaneousFrequencyEstimation2024}. In the weak-signal regime, the lowest-frequency component $f_{\rm{min}} = \min_{m \in \mathbb{Z}} \bigl| \delta f_{\rm{sig}} - m f_{\rm{mod}} \bigr|$ observed in the output spectrum and the frequency offset $\delta f_{\rm{sig}} = |f_{\text{sig}} - f_c|$ satisfy the congruence relation
    \begin{equation}
        f_{\rm{min}} \equiv \pm \delta f_{\rm{sig}} \pmod{f_{\rm{mod}}}
    \end{equation}
    since measuring $f_{\rm{min}}$ determines only the deviation from $\delta f_{\rm{sig}}$ to the nearest multiple of $f_{\rm{mod}}$, leaving the sign of the residue ambiguous. Consequently, a single measurement yields a solution space 
    \begin{equation}
        \mathcal{S} = \bigl\{ f_{\rm{min}} + m f_{\rm{mod}} \bigr\} \cup \bigl\{ (f_{\rm{mod}} - f_{\rm{min}}) + m f_{\rm{mod}} \bigr\},
    \end{equation}
    leaving $\delta f_{\rm{sig}}$ ambiguous up to the choice of sign and the integer $m~(m \in \mathbb{N})$.
    
    To resolve this ambiguity, a second measurement should be performed with a different modulation rate $f_{\rm{mod}}'$, yielding a distinct lowest-frequency component $f_{\text{min}}'$. The deviation $\delta f_{\rm{sig}}$ must then simultaneously satisfy
    \begin{equation}
        \begin{cases}
            \delta f_{\rm{sig}} \equiv \pm f_{\rm{min}} \pmod{f_{\rm{mod}}}, \\[4pt]
            \delta f_{\rm{sig}} \equiv \pm f_{\rm{min}}' \pmod{f_{\rm{mod}}'}.
        \end{cases}
    \end{equation}
    According to the Chinese Remainder Theorem, for each sign combination, if $\gcd(f_{\text{mod}}, f_{\text{mod}}') = 1$, this system has a unique solution in $[0, f_{\text{mod}} f_{\text{mod}}')$. Otherwise, the unambiguous range is reduced to $[0, \operatorname{lcm}(f_{\rm{mod}}, f_{\rm{mod}}'))$.
    
    However, the above derivation assumes ideal measurements. Measurement noise and finite spectral resolution practically introduce uncertainties in the extracted values of $f_{\rm{min}}$ and $f_{\rm{min}}'$, so that the congruence relations cannot precisely hold. We therefore recast the problem as a robust optimization. Defining the wrap-around distance on the modular circle as
        $d(\delta f_{\rm{sig}}; f_{\rm{min}}, f_{\rm{mod}}) = \min\Bigl\{ |\delta f_{\rm{sig}} \bmod f_{\rm{mod}} - f_{\rm{min}}|,\; f_{\rm{mod}} - |\delta f_{\rm{sig}} \bmod f_{\rm{mod}} - f_{\rm{min}}| \Bigr\}$,
    the optimal estimate minimizes the sum of squared distances over all sign combinations:
    \begin{equation}
        \widehat{\delta f}_{\rm{sig}} = \arg\min_{\delta f_{\rm{sig}}, s_1, s_2}\sum_{i=1}^{2} d^{2}\bigl(\delta f_{\rm{sig}}; s_i f_{\rm{min}}^{(i)}, f_{\rm{mod}}^{(i)}\bigr),
    \end{equation}
    where $\delta f_{\rm{sig}} \in [0, \Delta_{\max}]$, $\Delta_{\max} = \operatorname{lcm}\bigl(f_{\rm{mod}}, f_{\rm{mod}}'\bigr)$ defines the unambiguous estimation bandwidth and $s_1, s_2 = \pm 1$ account for the sign ambiguity in each measurement. This approach yields a precise estimate of $\delta f_{\rm{sig}}$ even in the presence of measurement uncertainties, thereby enabling unambiguous broadband frequency estimation.

    \textbf{Spectrum Sensing Enhancement}
    For a single-frequency signal field, we can achieve broadband spectrum estimation of the signal with the aid of the FMLO, where the spectrum estimation accuracy depends on the measurement precision of the IF frequency and the magnitude of frequency noise inherent in the signal field. In practice, any response peak in the IF spectrum can be used to estimate the signal frequency. We assume that multiple measurements can be performed using different frequency peaks identified in the spectrum, and the measured value $y_i$ can be expressed as
    \begin{equation}
        y_i = |f_{\rm{sig}} - f_{{\rm{LO}},i}| + \varepsilon_i,
    \end{equation}
    where $f_{{\rm{LO}},i}$ is the LO frequency component that mixes with $f_{\rm{sig}}$ to produce the corresponding response peak in the $i$-th measurement. $\varepsilon_i$ is the random error of the $i$-th measurement, which we assume follows a normal distribution $\varepsilon_i \sim \mathcal{N}(0,\sigma^{2})$. Therefore, for $m$ independent measurements, the best unbiased estimate of the signal frequency is
    \begin{equation}
        \hat{f}_{\rm{sig}} = \frac{1}{m} \sum_{i=1}^{m} \bigl( s_i y_i + f_{{\rm{LO}},i} \bigr),
    \end{equation}
    where $s_i$ takes the value $+1$ or $-1$ depending on the $i$-th measurement result $y_i$. In this case, the measurement of $f_{\rm{sig}}$ achieves a maximum precision of $\sigma / \sqrt{m}$. However, when the signal field is sufficiently strong, the IF response will contain higher-order mixing products. If the $m$ measurements include these higher-order frequency peaks, i.e., $y_i = |n_{i} f_{\rm{sig}} - f_{{\rm{LO}},i}| + \varepsilon_i$, then the Fisher information can be written as
    \begin{equation}
        I(f_{\rm{sig}}) = \sum_{i=1}^{m} \frac{1}{\sigma^2} \left( \frac{\partial y_i}{\partial f_{\rm{sig}}} \right)^{2} 
        = \frac{1}{\sigma^2} \sum_{i=1}^{m} n_i^{2},
        \label{eq:fisher_info}
    \end{equation}
    According to the Cram\'{e}r-Rao lower bound, the minimum possible variance of the estimate of $f_{\rm{sig}}$ is
    \begin{equation}
        \sigma_{\rm{est}} \geq \frac{\sigma}{\sqrt{\sum_{i=1}^{m} n_i^2}}.
    \end{equation}
    Therefore, exploiting higher-order frequency response components can, in principle, yield higher precision in frequency sensing.

\section*{Acknowledgements}
    The authors appreciate the instructive discussions with Prof. Hua-Dong Cheng from the Chinese Academy of Sciences. We acknowledge funding from the National Natural Science Foundation of China (Grant Nos. T2495253, 62435018,  12274059, 12574528, 1251101297 and W2541020), the National Key R and D Program of China (Grant No. 2022YFA1404002).

\section*{Data Availability}
    All experimental data used in this study are available from the corresponding author upon request.
    
\section*{CODE AVAILABILITY}
    The custom codes used to produce the results presented in this paper are available from the corresponding authors upon request.

\section*{Author contributions statement}
    B.-B.W. and  D.-S.D. conceived the idea. J.R.C conducted the physical experiments. The research was supervised by B.-B.W and D.-S.D. All authors contributed to discussions regarding the results and the analysis contained in the manuscript.

\section*{Competing interests}
    The authors declare no competing interests.

\bibliography{paper/ref}

\end{document}


\title{Supplementary Material: Compressive Spectrum Sensing via Spectral Multiplexing in Rydberg Atomic Receiver}

\author{Jun-Rong Chen$^{1,\textcolor{blue}{\star}}$}
\author{Yi-Ming Yin$^{2,3,\textcolor{blue}{\star}}$}
\author{Le-Bin Chen$^{6,\textcolor{blue}{\star}}$}
\author{Kai Wang$^{4,\textcolor{blue}{\star}}$}
\author{Bang Liu$^{2,3}$}
\author{Li-Hua Zhang$^{2,3}$}
\author{Hao Tian$^{1,\textcolor{blue}{\P}}$}
\author{Ming-Min Zhao$^{6,\textcolor{blue}{\S}}$}
\author{Bin-Bin Wei$^{5,\textcolor{blue}{\ddagger}}$}
\author{Dong-Sheng Ding$^{2,3,\textcolor{blue}{\dagger}}$}

\affiliation{$^1$School of physics, Harbin Institute of Technology, Harbin, Heilongjiang 150001, China.}
\affiliation{$^2$Key Laboratory of Quantum Information, University of Science and Technology of China, Hefei, Anhui 230026, China.}
\affiliation{$^3$Anhui Province Key Laboratory of Quantum Network, University of Science and Technology of China, Hefei, Anhui 230026, China.}
\affiliation{$^4$Electronic Engineering Institute, National University of Defense Technology, Hefei, Anhui 230037, China}
\affiliation{$^5$Institute of System Engineering, Tianjin 300161, China.}
\affiliation{$^6$College of Information Science and Electronic Engineering, Zhejiang University, Hangzhou, Zhejiang 310058, China}

\date{\today}

\symbolfootnote[1]{J.R.C., Y.M.Y., L.B.C. and K.W. contribute equally to this work.}
\symbolfootnote[2]{dds@ustc.edu.cn}
\symbolfootnote[3]{weibb.2009@tsinghua.org.cn}
\symbolfootnote[4]{zmmblack@zju.edu.cn}
\symbolfootnote[5]{tianhao@hit.edu.cn}

\maketitle

    In this Supplementary Material, we provide a detailed analysis of the physics-induced sensing matrix, its frequency-dependent response, and its role in enabling compressive spectral reconstruction and communication performance enhancement.

    \subsection{Properties of the Physics-Induced Sensing Matrix}
    
    Following the linear measurement model of compressive sensing, the measurement process can be formally expressed as $\mathbf{y} = \mathbf{A}\mathbf{x} + \mathbf{n}$, where $\mathbf{x}$ denotes the input spectrum and $\mathbf{y}$ represents the measured intermediate-frequency (IF) response \cite{choiCompressedSensingWireless2017, baydounCFARCompressedDetection2024, josephLowresolutionCompressedSensing2025, chenJointLocationSensing2024, IntegrationDeepLearning2025}. In this section, we present the numerical form of the sensing matrix $\mathbf{A}$ to explicitly illustrate how the measurement process maps input spectral components to IF responses. As discussed in the main text, the sensing matrix is directly induced by the multi-heterodyne mixing between the frequency-modulated local oscillator (FMLO) and the input signal. Each column of $\mathbf{A}$ represents the measurement signature of a single-frequency input, distributed across multiple IF channels according to $f_{\rm IF} = |\delta f_{\rm sig} - m f_{\rm mod}|$, with amplitudes governed by the Bessel envelope of the FMLO.

\begin{figure}
        \centering
        \includegraphics[width=1\linewidth]{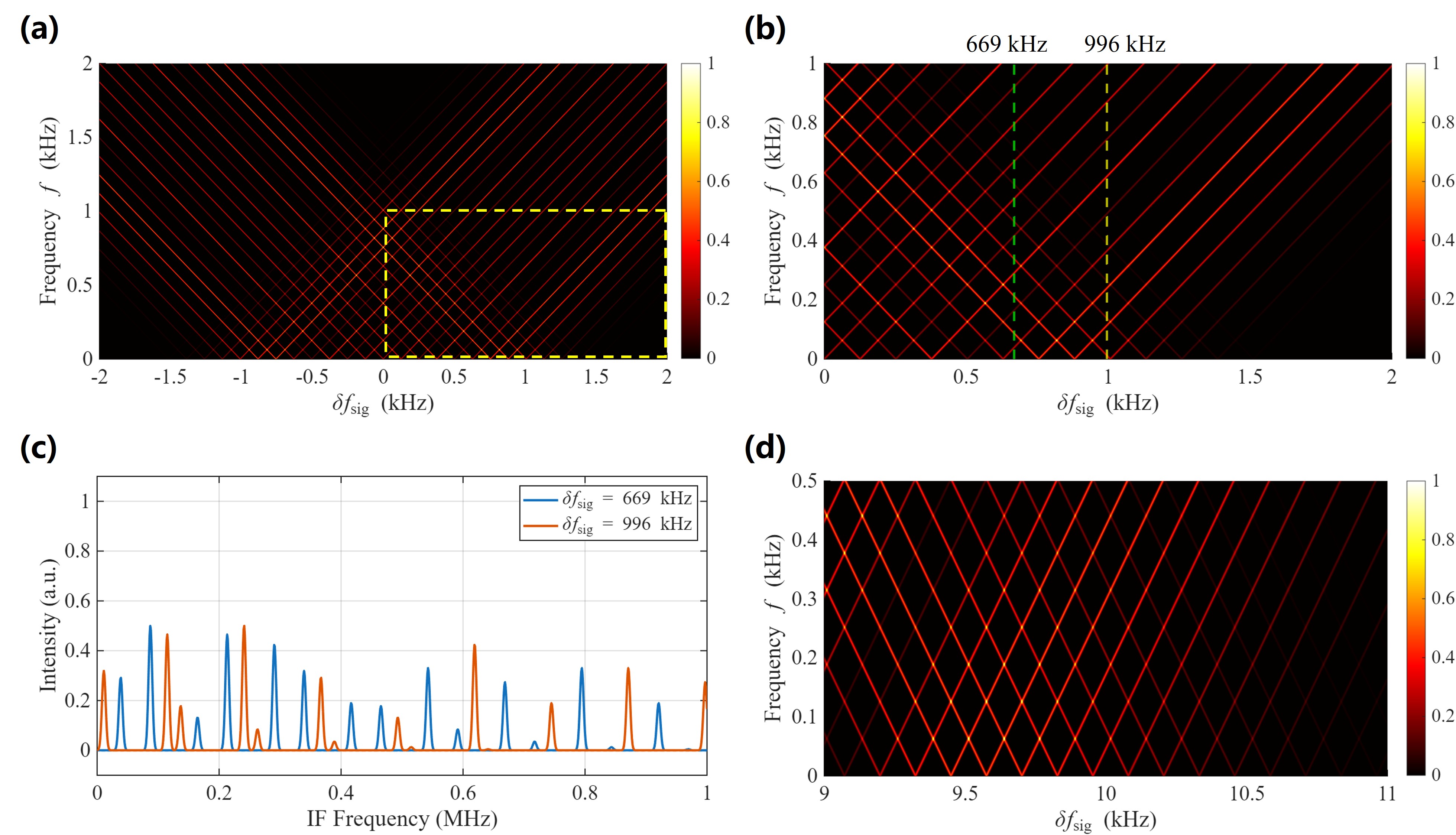}
        \caption{\textbf{Structure and interpretation of the physics-induced sensing matrix.} 
        (a) Simulated sensing matrix $\mathbf{A}$ arising from the multi-heterodyne mixing between the FMLO and the input signal. Each column represents the distributed IF response of a single-frequency input, with the structure governed by $f_{\rm IF} = |\delta f_{\rm sig} - m f_{\rm mod}|$ and weighted by the Bessel envelope. The yellow dashed box indicates the region shown in (b). 
        (b) Local view of the sensing matrix for $f_{\rm dev} = 1$~MHz. The green and yellow dashed lines correspond to $\delta f_{\rm sig} = 669~\mathrm{kHz}$ and $996~\mathrm{kHz}$, respectively. 
        (c) Cross-sectional profiles at the frequencies indicated in (b). The distinct response patterns demonstrate that different spectral components produce distinguishable measurement signatures across the IF channels.
        (d) Local view of the sensing matrix for $f_{\rm dev} = 10$~MHz in the range $\delta f_{\rm sig} \in [9, 11]$~MHz. The structure illustrates the sensing matrix behavior near the boundary of the FMLO spectral envelope, where the channel responses gradually diminish.}
        \label{SMfigA}
    \end{figure}

    Fig.~\ref{SMfigA} (a) shows the full sensing matrix, where a structured pattern emerges from the underlying frequency-mixing process, and the highlighted region is shown in detail in Fig.~\ref{SMfigA} (b) for an FMLO with $f_{\rm dev} = 1$~MHz. Fig.~\ref{SMfigA} (c) shows representative cross sections of the sensing matrix for selected input frequencies, taken at the positions indicated by the green and yellow dashed lines in Fig.~\ref{SMfigA} (b). The distinct response profiles demonstrate that different spectral components produce distinguishable measurement signatures across the IF channels, forming the basis for spectral reconstruction and establishing the sensing matrix as a physically structured measurement operator. To connect the numerical model with the experimental data presented in the main text, Fig.~\ref{SMfigA} (d) shows a local view of the sensing matrix constructed for $f_{\rm dev} = 10$~MHz, corresponding to the FMLO configuration used in Fig.~3 (a) of the main text. The structure observed in Fig.~\ref{SMfigA} (d) closely resembles the pattern measured experimentally, indicating that the essential features of the measurement process are well captured by the present model. It should be noted that the numerical construction of $\mathbf{A}$ does not include measurement noise or the frequency-dependent response of the Rydberg atoms. In the experiment, the atomic response decreases at higher frequencies, leading to weaker IF signal amplitudes in the corresponding region of Fig.~3 (a). In contrast, the simulated sensing matrix is solely determined by the FMLO spectrum, and thus the amplitude variation across frequency is governed only by the Bessel envelope. Despite these simplifications, the calculated matrix successfully reproduces the characteristic response patterns observed in the experiment. 
    
    The deterministic structure of the sensing matrix, evident in Figs.~\ref{SMfigA} (b) and (d), distinguishes the present approach from conventional compressive sensing. While the present framework shares similarities with compressive sensing, the sensing matrix is not explicitly engineered to satisfy strict mathematical conditions such as randomness or restricted isometry \cite{xiaDeterministicRandomMeasurement2014, shenCompressiveSensingShortestSolution}. Instead, it arises directly from the underlying physical mixing process. In contrast to conventional compressive sensing frameworks that rely on randomness, the sensing matrix here is fully deterministic and physically defined by the FMLO spectrum. Stable reconstruction is enabled by the fact that different spectral components produce distinguishable IF response patterns across the multiplexed channels, with only limited overlap. This physical separability of the measurement signatures provides the basis for reliable spectral recovery.

    \subsection{Frequency-Dependent Response of the Multiplexed Channels}

\begin{figure*}
        \centering
        \includegraphics[width=1\linewidth]{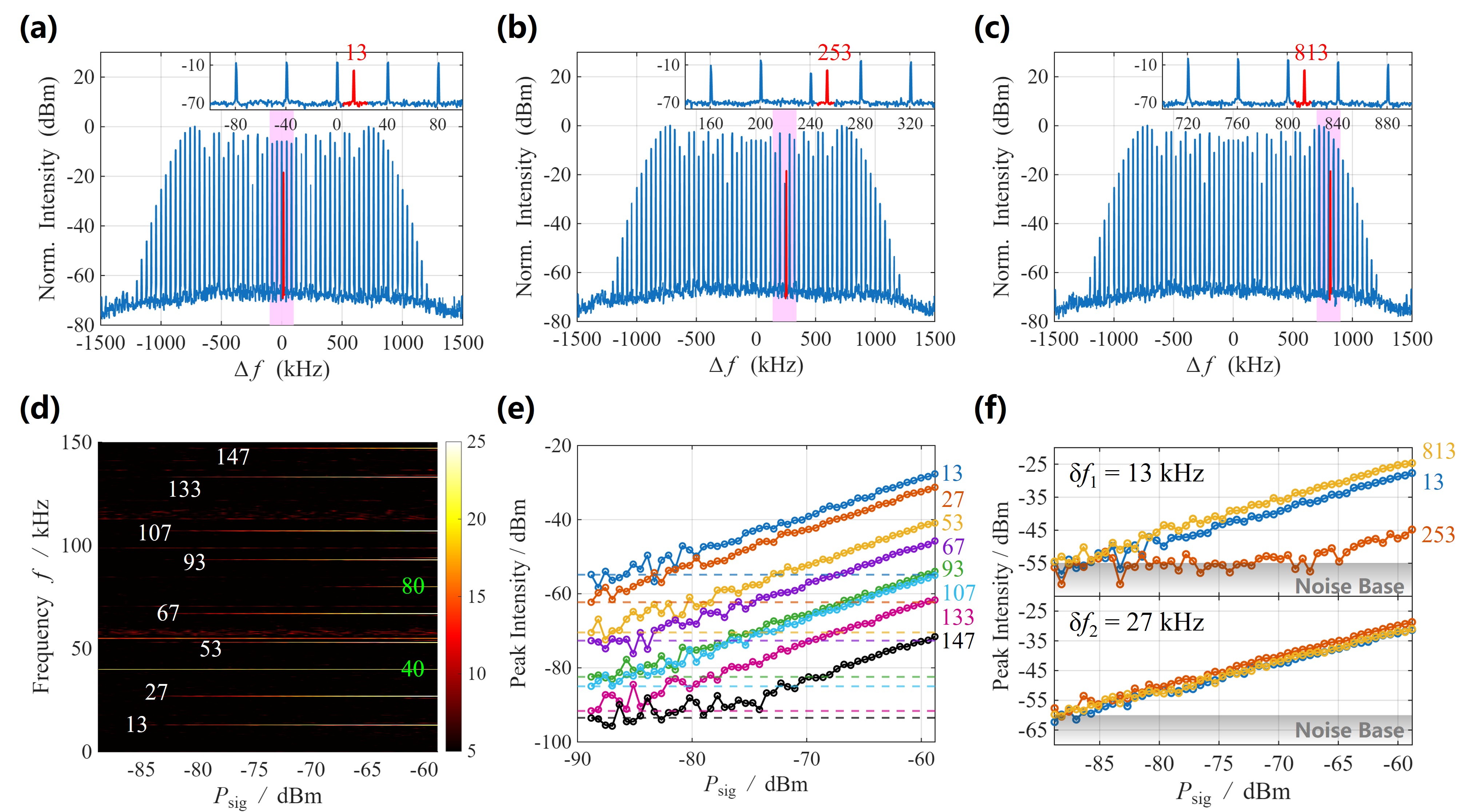}
        \caption{\textbf{Influence of the sensing matrix envelope on the compressive projections.} 
        The frequency deviation $f_{\rm{dev}}$ and modulation rate $f_{\rm{mod}}$ are fixed at 800 kHz and 40 kHz, respectively. Owing to the non-uniform intensity distribution among the rows of the sensing matrix determined by the Bessel envelope, the compressive projections vary for signal fields with different relative frequency detunings $\delta f_{\rm{sig}}$. Examples are shown for (a) $\delta f_{\rm{sig}} = 13$~kHz, (b) $\delta f_{\rm{sig}} = 253$~kHz and (c) $\delta f_{\rm{sig}} = 813$~kHz, where in all three cases the frequency interval to the nearest lower-frequency sensing matrix row is $\delta f_1 = 13$~kHz. 
        (d) Output compressive IF spectra as a function of the calibrated signal power $P_{\rm{sig}}$ for the case shown in (a) with $\delta f_{\rm{sig}} = 13$~kHz. 
        (e) Corresponding projection amplitudes at various frequencies versus $P_{\rm{sig}}$, with numbers adjacent to the data points indicating frequencies in kHz and horizontal dashed lines denoting the frequency-specific noise floor. 
        (f) Response curves of the projection amplitudes at frequencies $f = \delta f_1$ and $f = \delta f_2$ versus $P_{\rm{sig}}$ for the three cases in (a)-(c). Numbers adjacent to the data points represent the corresponding signal frequency deviations in kHz.}
        \label{SMfig1}
    \end{figure*}

    The compressive spectral multiplexing model described in the main text establishes that each frequency component of the FMLO serves as an independent down-conversion channel, corresponding to a row of the sensing matrix $\mathbf{A}$ with a weight determined by the Bessel envelope $J_m(\beta)$. A direct consequence of this structure is that the compressive projection of an external signal field depends not only on the signal frequency but also on the specific sensing matrix rows involved in the multi-heterodyne mixing process. This dependence is systematically characterized in Fig.~\ref{SMfig1} for an FMLO with $f_{\rm{dev}} = 800$~kHz and $f_{\rm{mod}} = 40$~kHz.
    
    Fig.~\ref{SMfig1} (a)-(c) show the spectral arrangement for three single-frequency signal fields with different detunings relative to the FMLO center. In Fig.~\ref{SMfig1}(a), the signal at $\delta f_{\rm{sig}} = 13$~kHz lies near the central region of the FMLO spectrum, where the sensing matrix rows have strong Bessel coefficients. In Fig.~\ref{SMfig1}(b), the signal at $\delta f_{\rm{sig}} = 253$~kHz is positioned adjacent to the FMLO component at $240$~kHz, whose intensity is considerably weaker than that of the central rows. In Fig.~\ref{SMfig1}(c), the signal at $\delta f_{\rm{sig}} = 813$~kHz is located near the edge of the FMLO spectrum. For a frequency-modulated signal, the FMLO components in the vicinity of $\pm f_{\rm{dev}}$ generally retain appreciable intensity as an intrinsic property of the Bessel envelope, and therefore the signal in this case still undergoes compressive projections with non-negligible weights. In all three cases, the frequency separation to the nearest lower-frequency FMLO component is $\delta f_1 = 13$~kHz.
    
    Fig.~\ref{SMfig1}(d) presents the IF output spectra for the configuration of Fig.~\ref{SMfig1}(a) as the calibrated signal power $P_{\rm{sig}}$ is gradually reduced from $-60$~dBm. The spectra exhibit multiple response peaks at frequencies of $13$, $27$, $53$, $67$, $93$, $107$, $133$, and $147$~kHz, each arising from the signal undergoing a compressive projection onto a different row of the sensing matrix. As $P_{\rm{sig}}$ decreases, the amplitudes of these projections decay at different rates. The peak amplitudes extracted from the spectra are plotted against $P_{\rm{sig}}$ in Fig.~\ref{SMfig1}(e). Owing to shot noise, low-frequency noise, and other noise contributions, the noise floor is frequency-dependent. The noise floor specific to each peak is indicated by a horizontal dashed line of the corresponding color. Despite these differences in the absolute noise level, all peaks descend into their respective noise floors at comparable signal powers, approximately in the range of $-80$~dBm to $-90$~dBm. The variations among the decay rates can be attributed to the different weights of the sensing matrix rows involved in the respective projections. For instance, the peak at $147$~kHz decays most rapidly as it originates from the projection onto the FMLO component at $160$~kHz, whose sensing matrix weight is approximately $8$~dB weaker than that of the central row at $0$~kHz. Consequently, this peak exhibits a considerably faster decline with decreasing $P_{\rm{sig}}$ than the peak at $13$~kHz, which involves the strongest central row of the sensing matrix. Fig.~\ref{SMfig1}(f) compares the power-dependent response of the two principal IF peaks at $f = \delta f_1$ and $f = \delta f_2$ across the three configurations of Figs.~\ref{SMfig1}(a)-(c), with the three cases distinguished by blue, orange, and yellow curves. The most notable feature appears in the $\delta f_1$ peak for Fig.~\ref{SMfig1}(b) (orange, upper panel). Here the nearest lower-frequency sensing matrix row at $240$~kHz is approximately $17$~dB weaker than the corresponding rows in the other two configurations, and the $\delta f_1$ projection is correspondingly weaker by the same margin, descending into the noise floor at a higher signal power. In contrast, the $\delta f_2$ peaks in the lower panel are nearly identical across all three cases, as the nearest higher-frequency sensing matrix row has comparable and strong weight in each configuration. These observations confirm that each IF projection is governed by the weight of the corresponding sensing matrix row, directly motivating the normalization procedure used in Fig.~4 of the main text, where the measured IF amplitudes are compensated by the known sensing matrix envelope to account for the non-uniform channel sensitivity.

    \begin{figure*}
        \centering
        \includegraphics[width=1\linewidth]{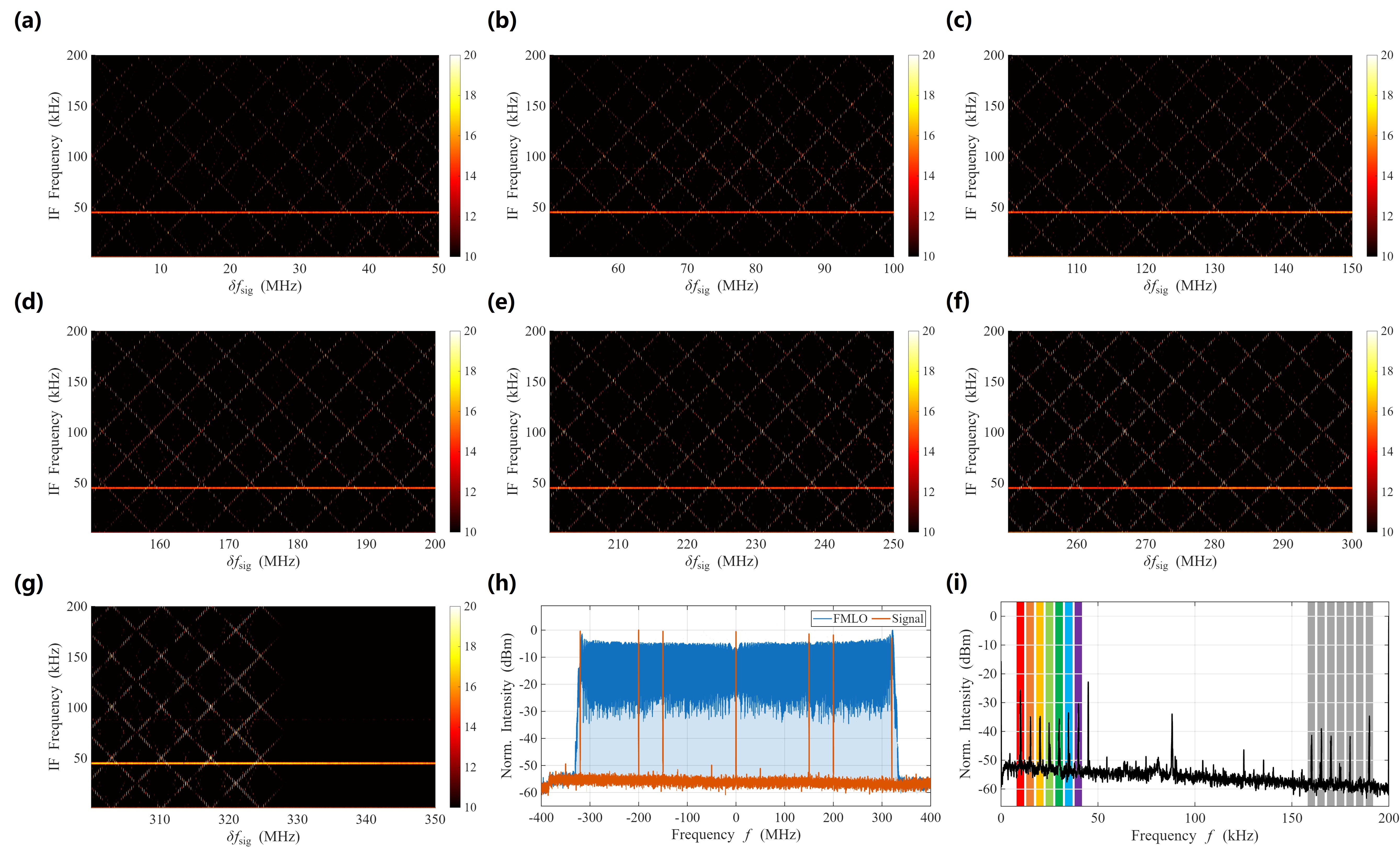}
        \caption{\textbf{Extended recoverable spectral range with a wideband FMLO configuration.} 
        (a)-(g) Compressive IF spectra of the system under an FMLO with $f_{\rm{dev}} = 320$~MHz and $f_{\rm{mod}} = 252$~kHz, measured over the signal frequency range of 0-350~MHz in seven contiguous 50~MHz segments. The system produces a detectable response for signal frequencies up to approximately 326~MHz, corresponding to a single-sideband recoverable spectral range of $B_{\rm{recover}} \approx 326$~MHz. The response amplitude varies with signal frequency due to the nonuniform Bessel envelope of the FMLO, yet a compressive projection is obtained at every frequency within this range. 
        (h) Spectral arrangement of a seven-tone signal field and the FMLO with $f_{\rm{dev}} = 320$~MHz and $f_{\rm{mod}} = 200$~kHz. The signal contains frequency components at offsets of $-320$~MHz~$+$~10~kHz, $-200$~MHz~$+$~15~kHz, $-150$~MHz~$+$~20~kHz, $0$~MHz~$+$~25~kHz, $+150$~MHz~$+$~30~kHz, $+200$~MHz~$+$~35~kHz, and $+320$~MHz~$+$~40~kHz relative to the FMLO center. (i) Composite IF spectrum resulting from the compressive multi-heterodyne mixing of the seven-tone signal with the FMLO. The seven rainbow-colored shaded regions mark the direct IF outputs corresponding to the individual signal components, and gray shaded regions mark additional IF replicas generated by the multi-heterodyne mapping.}
        \label{SMfig2}
    \end{figure*}

    \subsection{Extension of the Recoverable Spectral Range Using a Wideband FMLO}

    As established in the main text, the recoverable spectral range $B_{\rm{recover}}$ scales with the FMLO frequency deviation $f_{\rm{dev}}$ and is independent of the intrinsic atomic instantaneous bandwidth. To demonstrate this scalability, we extend $f_{\rm{dev}}$ to 320~MHz while maintaining $f_{\rm{mod}} = 252$~kHz. The FMLO in this configuration is generated by a Rohde \& Schwarz SBA100B analog signal generator. Figures~\ref{SMfig2} (a)-(g) present the compressive IF spectra measured over the signal frequency range of 0-350~MHz in contiguous 50~MHz segments. A detectable response is observed up to approximately 326~MHz, confirming a single-sideband recoverable spectral range of $B_{\rm{recover}} \approx 326$~MHz. This observation is consistent with the approximate scaling relation $B_{\rm{recover}} \propto f_{\rm{dev}}$, as discussed in the main text. The response amplitude varies with signal frequency owing to the nonuniform Bessel envelope of the FMLO, yet a compressive projection is obtained at every frequency within this range. It should be noted that a larger $f_{\rm{dev}}$ increases the number of FMLO frequency components for a fixed $f_{\rm{mod}}$ and total LO power, thereby reducing the power allocated to each individual tooth. This power reduction directly weakens the atomic response per channel, leading to a reduction in the effective SNR of each sensing branch. This reveals a fundamental trade-off between bandwidth expansion and per-channel sensitivity. In the limit of very large $f_{\rm{dev}}$, the power per FMLO tooth becomes progressively reduced, leading to weaker projections and lower SNR for individual sensing channels. This degradation in per-channel sensitivity can ultimately limit the reconstruction fidelity.
    
    To illustrate the capability of multi-frequency signal recovery under this wideband configuration, we employ a seven-tone signal with frequency components distributed across the full $\pm 320$~MHz range, as depicted in Fig.~\ref{SMfig2} (h). For this measurement, $f_{\rm{mod}}$ is set to 200~kHz to ensure clear spectral separation of the IF outputs. Owing to the large ratio of the total FMLO span (640~MHz) to the tone spacing and the 2~kHz resolution bandwidth of the spectrum analyzer, the discrete FMLO spectrum appears as a quasi-continuous envelope, shown as the blue shaded region with boundary indicated by the blue solid line. The resulting composite IF spectrum is shown in Fig.~\ref{SMfig2} (i). All seven signal components are clearly resolved as distinct IF outputs, each corresponding to a compressive projection onto a different sensing matrix row. Additional IF replicas generated by the multi-heterodyne mapping are also present in the IF spectrum.

    \subsection{BPSK Signal Processing and Multi-Branch Maximal-Ratio Combining}
    
    In our experiment, the BPSK transmission chain begins with bit-to-symbol mapping. Each transmitted bit $b_k \in \{0, 1\}$ is mapped to a polar symbol $a_k \in \{+1, -1\}$ according to $a_k = 1 - 2b_k$. Prior to up-conversion, the discrete symbol sequence is upsampled and passed through a root-raised-cosine pulse-shaping filter to confine the signal within the allocated bandwidth and to satisfy the zero intersymbol interference condition at the receiver sampling instants. The shaped complex baseband waveform is
    \begin{equation}
        s(t) = \sum_{k} a_k \, p(t - kT),
        \label{eq_s1}
    \end{equation}
    where $p(t)$ is the shaping pulse and $T$ is the symbol interval. The real-valued transmitted signal at carrier frequency $f_c$ is $x(t) = \Re\{s(t) e^{\rm{i} 2\pi \it{f_c t}}\}$. In the multi-channel configuration of Fig.~5 in the main text, seven such BPSK branches are generated independently with distinct carrier frequencies and combined into a composite waveform before being radiated toward the Rydberg atomic receiver.
    
    During propagation and detection, the signal inevitably undergoes attenuation, phase rotation, and frequency offset, and additive noise is introduced at the receiver front end. To recover the transmitted symbols, we perform a sequence of digital signal processing steps on the photodetector output. Firstly, the sampled signal is digitally down-converted and low-pass filtered to extract the complex baseband envelope. A matched filter is then applied to maximize the signal-to-noise ratio (SNR) while preserving the pulse shape.
    
    To compensate for the time-varying channel effects introduced by the atomic response and the hardware chain, we perform timing recovery and frequency offset correction based on a preamble of 260 pseudo-random bits generated with a fixed random seed [rng(12345)]. Timing recovery determines the optimal sampling phase within each symbol interval by maximizing the correlation between the received preamble and the known reference sequence, ensuring that sampling occurs at the point of maximum eye opening. Frequency offset correction estimates the residual carrier offset $\Delta f$ from the linear phase trend across the preamble symbols, $\phi[n] \approx \phi_0 + 2\pi \Delta f n T$, and compensates it by multiplying the baseband signal with $\exp(-\rm{i} 2\pi \Delta \it{f t})$. After these corrections, a first-order complex gain equalizer is applied using the channel estimate $\hat{h}$ obtained from the preamble, and the received symbols are normalized as $z_k = y_k / \hat{h}$. Each communication frame carries 1500 information bits, and the bit error rate data presented in Fig.~5(e) in the main text are obtained from more than $1 \times 10^5$ transmitted bits.
    
    After equalization, soft demodulation is performed to compute the log-likelihood ratio (LLR) for each bit. Assuming the equalized symbols are normalized and the residual noise is approximately Gaussian, the LLR is proportional to the real part of the equalized symbol scaled by the inverse noise variance,
    \begin{equation}
        \mathrm{LLR}_k = \log \frac{p(z_k \mid a_k = +1)}{p(z_k \mid a_k = -1)} \propto \frac{2 \Re\{z_k\}}{\sigma^2},
        \label{eq_s2}
    \end{equation}
    where $\sigma^2$ is the equivalent noise variance of the demodulation branch. The sign of $\mathrm{LLR}_k$ determines the decoded bit, and its magnitude reflects the reliability of the decision.
    
    In our experiment, the same transmitted symbol is sensed through multiple IF replicas generated by the FMLO multiplexed channels. Each replica constitutes an independent diversity branch with its own SNR $\gamma_m = h_m^2 / \sigma_m^2$, as defined in Eq.~(7) of the main text. For the $m$-th replica, the corresponding LLR satisfies $\mathrm{LLR}_{k,m} \propto \gamma_m$, reflecting that branches with higher SNR produce more reliable soft decisions. When maximal-ratio combining (MRC) is applied, the LLRs from all available replicas are combined as a weighted sum
    \begin{equation}
        \mathrm{LLR}_k^{\mathrm{MRC}} = \sum_{m} w_m \, \mathrm{LLR}_{k,m}, \quad w_m \propto \gamma_m,
        \label{eq_s3}
    \end{equation}
    where each weight is proportional to the SNR of the corresponding branch. Since the effective SNR after combining obeys $\gamma_{\mathrm{MRC}} = \sum_m \gamma_m$, the MRC operation at the LLR level naturally realizes the SNR accumulation predicted by the compressive measurement model, where each replica corresponds to an independent noisy projection of the same transmitted symbol \cite{fengSurveyMaximumRatio2024}. Accordingly, FMLO-induced spectral replicas act as intrinsic diversity branches, such that maximal-ratio combining directly yields SNR accumulation without requiring explicit multi-antenna or multi-receiver architectures.

\bibliography{SM/SMref}